\documentclass[useAMS,usenatbib]{mn2e}
\usepackage{longtable}
\usepackage{graphicx}
\usepackage{lscape}

\newcommand{\apss}{ASS}

\newcommand{\Teff}{\mbox{$T_{\mathrm{eff}}$}}

\newcommand{\Porb}{\mbox{$P_{\mathrm{orb}}$}}

\newcommand{\Lines}[3]{\Ion{#1}{#2}\,$\lambda\lambda$\,#3}
\newcommand{\Ion}[2]{#1{\,\scriptsize #2}}

\newcommand{\kms}{\mbox{$\mathrm{km\,s^{-1}}$}}

\title[Post Common Envelope Binaries from SDSS]{Post Common Envelope
  Binaries from SDSS. I: 101 white dwarf main sequence binaries with
  multiple SDSS spectroscopy}

\author[A. Rebassa-Mansergas et al.]{
A. Rebassa-Mansergas$^1$, B. T. G\"ansicke$^1$,
P.Rodr\'{\i}guez-Gil$^{2}$, M.R. Schreiber$^3$, D. Koester$^4$\\
$^{1}$ Department of Physics, University of Warwick, Coventry CV4 7AL, UK \\
$^{2}$ Instituto de Astrof\'\i sica de Canarias, V\'\i a L\'actea, s/n, La Laguna, 
E-38205, Tenerife, Spain\\
$^{3}$ Departamento de Fisica y Astronomia, Universidad de Valparaiso, 
Avenida Gran Bretana 1111, Valparaiso, Chile \\
$^{4}$ Institut f\"ur Theoretische Physik und Astrophysik, University of Kiel,
24098 Kiel, Germany\\
}

\begin{document}
\date{Accepted 2007. Received 2007; in original form 2007}
\pagerange{\pageref{firstpage}--\pageref{lastpage}} \pubyear{2007}
\maketitle

\begin{abstract}
We present a detailed analysis of 101 white dwarf-main sequence
binaries (WDMS) from the Sloan Digital Sky Survey (SDSS) for which
multiple SDSS spectra are available. We detect significant radial
velocity variations in 18 WDMS, identifying them as post
common envelope binaries (PCEBs) or strong PCEB candidates. Strict
upper limits to the orbital periods are calculated, ranging from 0.43
to 7880\,d. Given the sparse temporal sampling and relatively low
spectral resolution of the SDSS spectra, our results imply a PCEB
fraction of $\ga$15\,\%  among the WDMS in the SDSS data
base.  Using a spectral decomposition/fitting technique we determined
the white dwarf effective temperatures and surface gravities, masses,
and secondary star spectral types for all WDMS in our sample. Two
independent distance estimates are obtained from the flux scaling
factors between the WDMS spectra, and the white dwarf models and main
sequence star templates, respectively. Approximately one third of the
systems in our sample show a significant discrepancy between the two
distance estimates.  In the majority of discrepant cases, the distance
estimate based on the secondary star is too large. A possible
explanation for this behaviour is that the secondary star spectral
types that we determined from the SDSS spectra are systematically too
early by 1--2 spectral classes. This behaviour could be explained by
stellar activity, if covering a significant fraction of the star by
cool dark spots will raise the temperature of the inter-spot regions.
Finally, we discuss the selection effects of the WDMS sample provided
by the SDSS project.
\end{abstract}


\label{firstpage}

\section{Introduction}
A large fraction of all stars in the sky are part of binary or
multiple systems \citep{iben91-1}. If the initial separation of the
main-sequence binary is small enough, the more massive star will
engulf its companion while evolving into a red giant, and the system
enters a common envelope (CE, e.g. \citealt{livio+soker88-1,
iben+livio93-1}). Friction within the CE leads to a rapid decrease of
the binary separation and orbital period, and the energy and angular
momentum extracted from the binary orbit eventually ejects the
CE. Products of CE evolution include a wide range of important
astronomical objects, such as e.g. high- and low-mass X-ray binaries,
double degenerate white dwarf and neutron star binaries, cataclysmic
variables and super-soft X-ray sources~--~with some of those objects
evolving at later stages into type Ia supernova and short gamma-ray
bursts.  While the concept of CE evolution is simple, its details are
poorly understood, and are typically described by  parametrised
models \citep{paczynski76-1, nelemansetal00-1, nelemans+tout05-1}. 
Consequently, population models of all types of CE
products are subject to substantial uncertainties.

Real progress in our understanding of close binary evolution is most
likely to arise from the analysis of post common envelope binaries
(PCEBs) that are both numerous and well-understood in terms of their
stellar components~--~such as PCEBs containing a white dwarf and a
main sequence star\footnote{Throughout this paper, we will use the
term WDMS to refer to the total class of white dwarf plus main
sequence binaries, and PCEBs to those WDMS that underwent a CE
phase.}. While detailed population models are already available,
\citep[e.g.][]{willems+kolb04-1}, there is a clear lack of
observational constraints. \citet{schreiber+gaensicke03-1} showed that
the sample of well-studied PCEBs is not only small, but being drawn
mainly from ``blue'' quasar surveys, it is also heavily biased towards
young systems with low-mass secondary stars~--~clearly not
representative of the intrinsic PCEB population.

\begin{figure}
\includegraphics[angle=-90,width=\columnwidth]{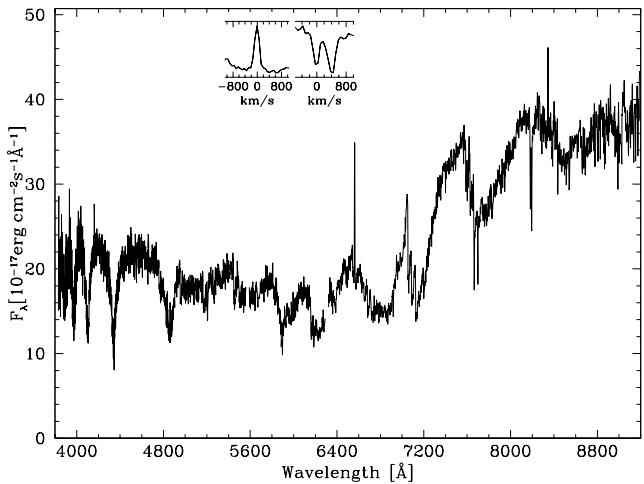}
\caption{The spectrum of SDSS\,J005208.42-005134.6, a typical WDMS in
the SDSS data base. The white dwarf is clearly visible in the blue
while the low mass companion dominates the red part of the
spectrum. Evident are the H$\alpha$ emission line, and the
\Lines{Na}{I}{8183.27,8194.81} absorption doublet, originating on the
companion star. These features are shown in the small insets on a
velocity scale, and are used to measure the radial velocities of 101
WDMS for which multiple SDSS spectra exist in DR5. See also
Fig.\,~\ref{f-fit}.}
\label{f-examp}
\end{figure}

The Sloan Digital Sky Survey (SDSS) is currently providing the
possibility of dramatically improving the observational side of PCEB
studies, as it has already identified close to 1000 WDMS (see 
Fig.\,\ref{f-examp}) with hundreds more to follow in future data releases
\citep{raymondetal03-1, silvestrietal06-1, eisensteinetal06-1,
southworthetal07-1}. Within SEGUE, a dedicated program to identify
WDMS containing cold white dwarfs is successfully underway
\citep{schreiberetal07-1}. 

Identifying all PCEBs among the SDSS WDMS, and determining their binary
parameters is a significant observational challenge. Here, we make
use of SDSS spectroscopic repeat observations to identify 18 PCEBs
and PCEB candidates from radial velocity variations, which are
excellent systems for in-depth follow-up studies. The structure of the
paper is as follows: we describe our WDMS sample and the methods used
to determine radial velocities in Sect.\,\ref{s-PCEB}. In
Sect.\,\ref{s-stellar} we determine the stellar parameters of the WDMS
in our sample. In Sect.\,\ref{s-discussion}, we discuss the fraction
of PCEBs found, the distribution of stellar parameters, compare our
results to those of \citet{raymondetal03-1} and
\citet{silvestrietal06-1}, discuss the incidence of stellar activity
on the secondary stars in WDMS, and outline the selection effects of
SDSS regarding WDMS with different types of stellar components.

\section{Identifying PCEBs in SDSS}
\label{s-PCEB}

SDSS operates a custom-built 2.5\,m telescope at Apache Point
Observatory, New Mexico, to obtain $ugriz$ imaging with a
120-megapixel camera covering $1.5\,\mathrm{deg}^2$ at once. Based on
colours and morphology, objects are then flagged for spectroscopic
follow-up using a fibre-fed spectrograph. Each ``spectral plate''
refers physically to a metal plate with holes drilled at the positions
of 640 spectroscopic plus calibration targets, covering
$\sim7\mathrm{deg}^2$. Technical details on SDSS are given by
\citet{yorketal00-1} and \citet{stoughtonetal02-1}. The main aim of
SDSS is the identification of galaxies \citep[e.g.][]{straussetal02-1}
and quasars \citep[e.g.][]{adelman-mccarthyetal06-1}, with a small
number of fibres set aside for other projects, e.g. finding cataclysmic
variables and WDMS \citep{raymondetal03-1}.

\begin{figure}
\label{f-fit}
\centerline{\includegraphics[width=0.48\textwidth]{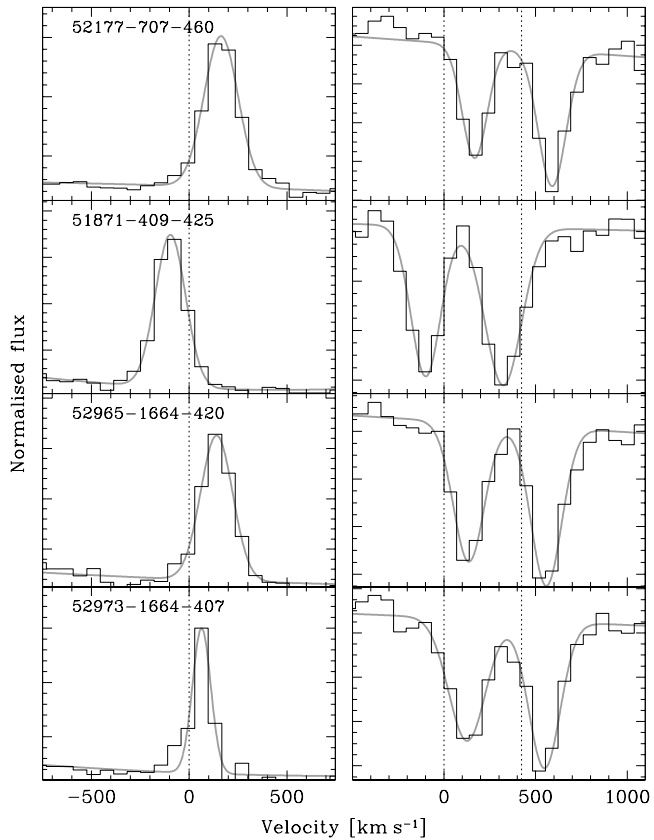}}
\caption{Fits to the \Lines{Na}{I}{8183.27,8194.81} absorption doublet
 (right panels) and the H$\alpha$ emission line (left panels) in the
 four SDSS spectra of the WDMS SDSS\,J024642.55+004137.2. The SDSS
 spectroscopic identifiers (MJD, Plate-ID and Fibre-ID) are given in
 the top left corner of the H$\alpha$ panels. \Ion{Na}{I} has been
 fitted with a double-Gaussian of fixed separation plus a parabola,
 H$\alpha$ with a Gaussian plus a parabola. In this system, radial
 velocity variations are already obvious to the eye. The top three
 spectra are taken in a single night, the bottom one is combined from
 data taken on three nights, MJD\,=\,52970, 52972, and 52973. The
 widths of the Gaussians fitting the \Ion{Na}{I} doublet are (top to
 bottom) 4.6\,\AA, 5.8\,\AA, 5.3\,\AA, and 6.0\,\AA.}
\end{figure}

A feature of SDSS hitherto unexplored in the study of WDMS is the fact
that $\sim10$ per cent of the spectroscopic SDSS objects are observed
more than once\footnote{SDSS occasionally re-observes entire spectral
plates, where all targets on that plate get an additional spectrum, or
has plates which overlap to some extent, so that a small subset of
targets on each plate is observed again.}: the detection of radial
velocity variations between different SDSS spectra of a given WDMS
will unambiguously identify such a system as a PCEB, or a strong PCEB
candidate. 
Throughout this paper, we define a PCEB as a WDMS with an
upper limit to its orbital period $\la300$\,d, a PCEB candidate as a
WDMS with periods $300\,\mathrm{d}\la
P_\mathrm{orb}\la1500\,\mathrm{d}$, following Fig.\,10 from
\citet{willems+kolb04-1}, which shows the period and mass distribution
of the present-day WDMS population at the start of the WDMS binary
phase. WDMS with period $\ga1500$\,d have too large binary separations
to undergo a CE phase, and remain wide systems. While these definitions
depend to some extend on the detailed configuration of the progenitor main
sequence binary, the population model of \citet{willems+kolb04-1}
predicts a rather clean dichotomy.

\begin{figure*}
\includegraphics[angle=-90,width=\textwidth]{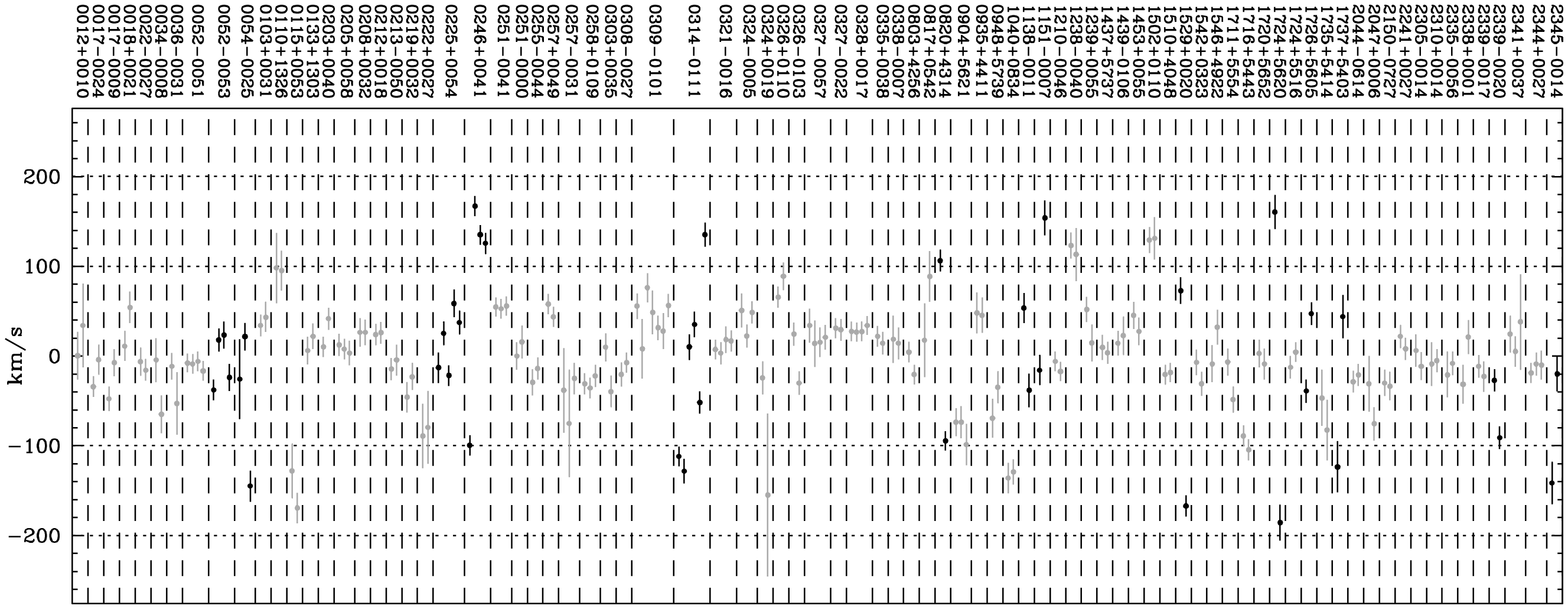}
\caption{\label{f-Na} Radial velocities obtained from the \Ion{Na}{I}
absorption doublet. WDMS $\ge3\sigma$ RVs variation,
i.e. PCEBs, are shown in black.}
\includegraphics[angle=-90,width=\textwidth]{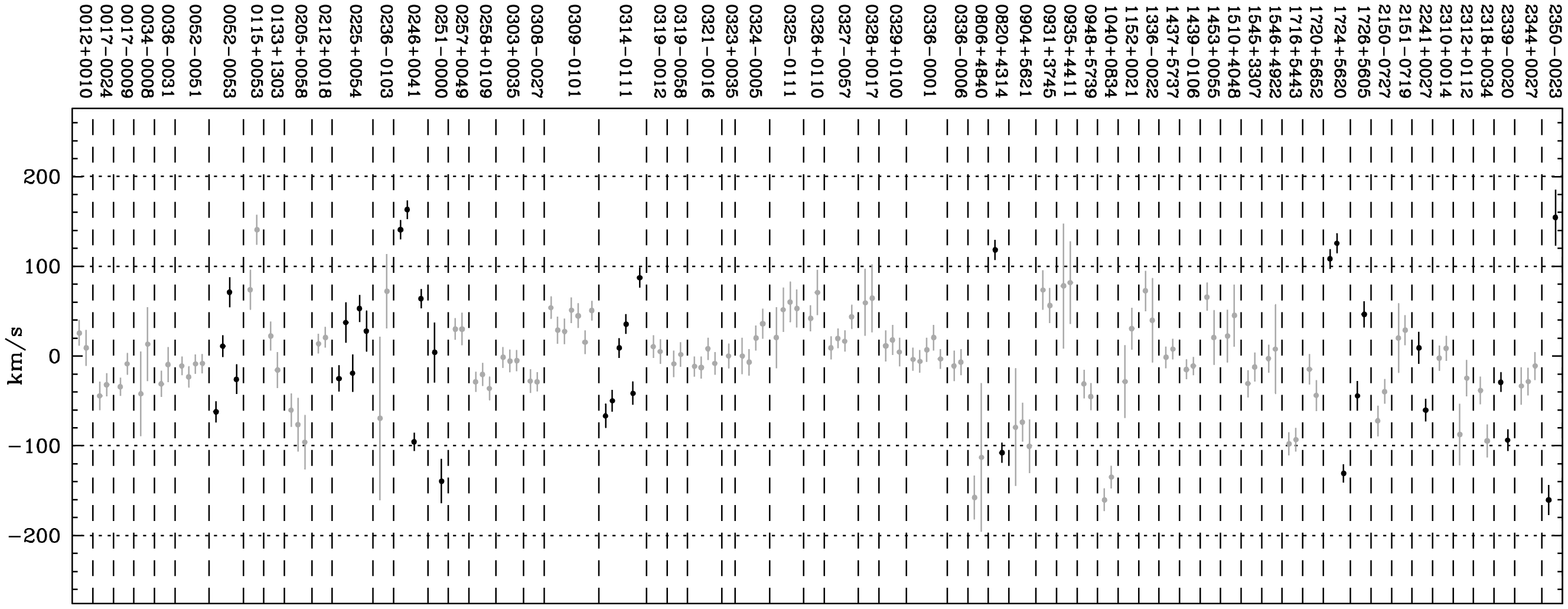}
\caption{\label{f-Ha} Same as Fig.~\ref{f-Na} but for the H$\alpha$ radial velocities.} 
\end{figure*}

We have searched the DR5 spectroscopic data base for multiple
exposures of all the WDMS listed by \citet{silvestrietal06-1} and
\citet{eisensteinetal06-1}, as well as a set of WDMS independently
found in the SDSS data by our team. This search resulted in a sample
of 130 WDMS with two to seven SDSS spectra. Among those WDMS, 101
systems have a clearly pronounced \Lines{Na}{I}{8183.27,8194.81}
absorption doublet and/or H$\alpha$ emission in their SDSS
spectra\footnote{The SDSS spectra are corrected to
heliocentric velocities and provided on a vacuum wavelengths scale.},
and were subjected to radial velocity measurements using one or both
spectral features. The \Ion{Na}{I} doublet was fitted with a second
order polynomial and double-Gaussian line profile of fixed
separation. Free parameters were the amplitude and the width of each
Gaussian and the velocity of the doublet. H$\alpha$ was fitted using a
second order polynomial plus a single Gaussian of free velocity,
amplitude and width. We computed the total error on the radial
velocities by quadratically adding the uncertainty in the zeropoint of
the SDSS wavelength calibration ($10\,km\,s^{-1}$,
\citealt{stoughtonetal02-1}) and the error in the position of the
\Ion{Na}{II}/H$\alpha$ lines determined from the Gaussian fits.
Figure\,\ref{f-fit} shows the fits to the four SDSS spectra of
SDSS\,J024642.55+004137.2, a WDMS displaying an extremely large radial
velocity variation identifying it as a definite PCEB. This figure also
illustrates an issue encountered for a handful of systems, i.e. that
the H$\alpha$ and \Ion{Na}{I} radial velocities do not agree in the
latest spectrum (Table\,\ref{t-rv}). This is probably related to the
inhomogeneous distribution of the H$\alpha$ emission over the surface
of the companion star, and will be discussed in more detail in
Sect.\,\ref{s-rv-porb}. In total, 18 WDMS show radial
velocity variations among their SDSS spectra at a $3 \sigma$ level and
qualify as PCEBs or strong PCEB candidates. Their radial velocities
are listed in Table\,\ref{t-rv} and illustrated in Fig.\,\ref{f-Na}
and Fig.\,\ref{f-Ha}. Three systems (SDSS0251--0000, SDSS1737+5403,
and SDSS2345--0014) are subject to systematic uncertainties in their
radial velocities due to the rather poor spectroscopic data. The
radial velocities for the remaining 83 WDMS  that did not show
any significant variation are available in the electronic edition of
the paper (see Table\,\ref{t-rvwdms}).

We note that special care needs to be taken in establishing the date
and time when the SDSS spectra where obtained: a significant fraction
of SDSS spectra are \textit{combined} from observations taken on
different nights (which we will call ``sub-spectra'' in what follows)
in which case the header keyword \texttt{MJDLIST} will be populated
with more than one date. The headers of the SDSS data provide the
exposure start and end times in International Atomic Time (TAI), and
refer to the start of the first spectrum, and the end of the last
spectrum. Hence, a meaningful time at mid-exposure can only be given
for those SDSS spectra that were obtained in a single contiguous
observation.

A crucial question is obviously how the fact that some of the spectra
in our sample are actually combinations of data from several nights
impacts our aim to identify PCEBs via radial velocity variations. To
answer this question, we first consider wide WDMS that did not undergo
a CE phase, i.e. binaries with orbital periods of
$\ga$\,years. For these systems, sub-spectra obtained over the course
of several days will show no significant radial velocity variation,
and combining them into a single spectrum will make no difference
except of increasing the total signal-to-noise ratio (S/N). In contrast to
this, for close binaries with periods of a few hours to a few days,
sub-spectra taken on different nights will sample different orbital
phases, and the combined SDSS spectrum will be a mean of those phases,
weighted by the S/N of the individual sub-spectra. In extreme cases,
e.g. sampling of the opposite quadrature phases, this may lead to 
smearing of the \Ion{Na}{I} doublet beyond recognition, or end up with
a very broad H$\alpha$ line. This may in fact explain the absence /
weakness of the \Ion{Na}{I} doublet in a number of WDMS where a
strong \Ion{Na}{I} doublet would be expected on the basis of the
spectral type of the companion. In most cases, however, the combined
SDSS spectrum will represent an ``effective'' orbital phase, and
comparing it to another SDSS spectrum, itself being combined or not,
still provides a measure of orbital motion. We conclude that the main
effect of the combined SDSS spectra is a decreased sensitivity to
radial velocity variations due to averaging out some orbital phase
information. Figure\,\ref{f-fit} shows an example of a combined
spectrum (bottom panel), which contains indeed the broadest
\Ion{Na}{I} lines among the four spectra of this WDMS.

In order to check the stability of the SDSS wavelength calibration
between repeat observations, we selected a total of 85 F-type stars
from the same spectral plates as our WDMS sample, and measured their
radial velocities from the \Lines{Ca}{II}{3933.67,3968.47} $H$ and $K$
doublet in an analogous fashion to the \Ion{Na}{I} measurement
carried out for the WDMS. None of those stars exhibited a significant
radial velocity variation, the maximum variation among all checked
F-stars had a statistical significance of $1.5\sigma$. The mean of the
radial velocity variations of these check stars was found to be
$14.5\,\mathrm{km\,s^{-1}}$, consistent with the claimed
$10\,\mathrm{km\,s^{-1}}$ accuracy of the zero-point of the wavelength
calibration for the spectra from an individual spectroscopic plate
\citep{stoughtonetal02-1}. In short, this test confirms that the SDSS
wavelength calibration is stable in time, and, as anticipated above
that averaging sub-spectra does not introduce any spurious radial
velocity shifts for sources that have no intrinsic radial velocity
variation (as the check stars are equally subject to the issue of
combining exposures from different nights into a single SDSS
spectrum). We are hence confident that any significant radial velocity
variation observed among the WDMS is intrinsic to the system.

\begin{table*}
\caption{\label{t-rv} Radial velocities of our 18 PCEBs and PCEB candidates,
measured from the H$\alpha$ emission line and/or the
\Lines{Na}{I}{8183.27,8194.81} absorption doublet. The HJDs for
SDSS spectra that have been combined from exposures taken in
several different nights (see Sect.~\ref{s-PCEB}) are set in italics.
PCEB candidates with uncertain radial velocity measurements are
indicated by colons preceding and trailing the object name. 
Upper limits of the orbital periods are also provided
(Sect.\,\ref{s-porb}). The two spectral components identified in the
spectra are coded as follows. DA\,=\,white dwarf with clearly visible
Balmer lines; DC\,=\,clearly visible blue continuum without noticeable
structure; blx\,=\,weak blue excess; dM\,=\,M-dwarf.}
\vspace*{-2ex}
\begin{minipage}[t]{0.47\textwidth}
\begin{center}
\setlength{\tabcolsep}{0.6ex}
\begin{tabular}{lcrclrclr}
\hline
SDSS\,J   & HJD & \multicolumn{3}{c}{RV(H$\alpha$)\,\kms} & \multicolumn{3}{c}{RV(Na)\,\kms} & $P_\mathrm{orb}\mathrm{[d]}<$\\
\hline
%
0052--0053   &  2451812.3463  &  71.3  &$\pm$&   16.6 &   23.4 &$\pm$&  14.9 & 280  \\
~~~~~DA/dM   &  \textit{2451872.6216}  &  11.0  &$\pm$&   12.0 &  18.0 &$\pm$& 12.7 &   \\
             &  \textit{2451907.0834}  & -62.1  &$\pm$&  11.7  &  -37.7&$\pm$& 11.6 &   \\
             &  2452201.3308  & -26.0  &$\pm$&   16.0 &  -23.8 &$\pm$&  14.9 &   \\
0054--0025   &  2451812.3463  &        &     &        &  21.6  &$\pm$&  15.3 &  4 \\
~~~~~DA/dM   &  \textit{2451872.6216}  &        &     &        &   -25.6&$\pm$&  44.2 &   \\
             &  \textit{2451907.0835}  &        &     &        & -144.7 &$\pm$&  17.2 &   \\
0225+0054    &  2451817.3966  &  53.0  &$\pm$&   14.9 &   58.6 &$\pm$&  15.4  & 45 \\
~~~~~~blx/dM &  2451869.2588  & -19.2  &$\pm$&   20.7 &  -21.6 &$\pm$&  11.6  &  \\
             &  2451876.2404  &  37.5  &$\pm$&   22.4 &   25.3 &$\pm$&  12.4  &  \\
             &  2451900.1605  & -25.1  &$\pm$&   14.0 &  -12.8 &$\pm$&  17.0  &  \\
             &  2452238.2698  &  27.9  &$\pm$&   22.3 &   37.4 &$\pm$&  12.8  &  \\
0246+0041    &  2451871.2731  & -95.5  &$\pm$&   10.2 &  -99.3 &$\pm$&  11.1  & 2.5 \\
~~~~~DA/dM   &  2452177.4531  &  163.1 &$\pm$&   10.3 &  167.2 &$\pm$&  11.3  &  \\
             &  2452965.2607  & 140.7  &$\pm$&   10.8 &  135.3 &$\pm$&  11.0  &  \\
             &  \textit{2452971.7468}  &  64.0  &$\pm$&    10.5 &  125.7 &$\pm$&  11.3 &   \\
:0251--0000: &  \textit{2452174.4732}  &  4.1   &$\pm$&   33.5 &    0.0 &$\pm$&  15.4 &  0.58 \\
~~~~~DA/dM   &  2452177.4530  & -139.3 &$\pm$&   24.6 &   15.8 &$\pm$&  18.3  &  \\
0309--0101   &  2451931.1241  &  44.8  &$\pm$&   13.2 &   31.5 &$\pm$&  13.7 & 153  \\
~~~~~DA/dM   &  2452203.4500  &  51.2  &$\pm$&   14.1 &   48.7 &$\pm$&  24.4 &   \\
             &  2452235.2865  &  27.4  &$\pm$&   14.1 &   76.3 &$\pm$&  16.2 &   \\
             &  2452250.2457  &  28.8  &$\pm$&   15.0 &    8.1 &$\pm$&  33.0 &   \\
             &  2452254.2052  &  53.9  &$\pm$&   11.8 &   55.7 &$\pm$&  14.3 &   \\
             &  2452258.2194  &  15.5  &$\pm$&   13.0 &   27.9 &$\pm$&  19.9 &   \\
             &  \textit{2453383.6493}  &  50.7  &$\pm$&   11.0 &   56.5 &$\pm$&  12.8 &   \\
0314--0111   &  2451931.1242  & -41.6  &$\pm$&   12.4 &  -51.7 &$\pm$&  12.4 & 1.1  \\
~~~~~DC/dM   &  2452202.3882  &  35.6  &$\pm$&   10.9 &   35.2 &$\pm$&  14.4 &   \\
\\
\hline
\end{tabular}
\end{center}
\end{minipage}
\hfill
\begin{minipage}[t]{0.47\textwidth}
\begin{center}
\setlength{\tabcolsep}{0.6ex}
\begin{tabular}{lcrclrclr}
\hline
SDSS\,J   & HJD & \multicolumn{3}{c}{RV(H$\alpha$)\,\kms} & \multicolumn{3}{c}{RV(Na)\,\kms} & $P_\mathrm{orb}\mathrm{[d]} <$\\
\hline
             &  2452235.2865  &   9.1  &$\pm$&   11.0 &   10.3 &$\pm$&  14.2 &   \\
             &  2452250.2457  & -49.8  &$\pm$&   12.2 & -128.2 &$\pm$&  13.9 &   \\
             &  2452254.2053  &  -66.7  &$\pm$&  12.8 & -111.7 &$\pm$&  10.9	&\\
             &  2452258.2195  &   87.3  &$\pm$&  10.8 &  135.2 &$\pm$&  13.5   &\\
0820+4314    &  2451959.3074  &  118.3  &$\pm$&   11.4 &  106.3 &$\pm$&  11.5  & 2.4\\
~~~~~DA/dM   &  \textit{2452206.9572}  & -107.8  &$\pm$&   11.2 & -94.6 &$\pm$&  10.8 &	\\
1138--0011   &  \textit{2451629.8523}  &         &     &       &   53.5 &$\pm$&  16.9&  35  \\
~~~~~DA/dM   &  2451658.2128  &         &     &       &  -38.1 &$\pm$&  18.6 &   \\
1151--0007   &  2451662.1689  &         &     &       &  -15.8 &$\pm$&  15.1 & 4.4  \\
~~~~~DA/dM   &  2451943.4208  &         &     &       &  154.0 &$\pm$&  19.5 &  \\
1529+0020    &  2451641.4617  &         &     &       &   73.0 &$\pm$&  14.8 & 0.96  \\
~~~~~DA/dM   &  2451989.4595  &         &     &       & -167.2 &$\pm$&  11.8  &  \\
1724+5620    &  \textit{2451812.6712}  &  125.6  &$\pm$&  10.2 &  160.6 &$\pm$&  18.4 & 0.43 \\
~~~~~DA/dM   &  2451818.1149  &  108.3  &$\pm$&   11.1 &     -  &$\pm$&  -  &   \\
             &  \textit{2451997.9806}  & -130.6  &$\pm$&  10.3 & -185.5 &$\pm$&  20.1&   \\
1726+5605    &  \textit{2451812.6712}  &  -44.3  &$\pm$&  16.7 &  -38.9 &$\pm$&  12.9 &   29\\
~~~~~DA/dM   &  \textit{2451993.9805}  &   46.6  &$\pm$&  14.6 &  47.3 &$\pm$&  12.5&	\\
:1737+5403:  &  2451816.1187  &         &     &       & -123.5 &$\pm$&  28.6 & 6.6\\
~~~~~DA/dM   &  2451999.4602  &         &     &       &   44.0 &$\pm$&  24.0 &  \\
2241+0027    &  2453261.2749  &    9.1  &$\pm$&  17.9 &   22.0 &$\pm$&  12.4 & 7880 \\
~~~~~DA/dM   &  2452201.1311  &   -60.3 &$\pm$&  12.7 &    8.1 &$\pm$&  12.2 &	\\
2339--0020   &  \textit{2453355.5822}  &  -29.2  &$\pm$&  10.4 &  -27.1 &$\pm$&  12.3 & 120 \\
~~~~~DA/dM   &  2452525.3539  &  -93.6  &$\pm$&  12.3 &  -90.1 &$\pm$&  12.7 &  \\
:2345-0014:  &  2452524.3379  &         &     &       & -141.5 &$\pm$&  22.9 & 9.5 \\
~~~~~DA/dM   &  \textit{2453357.5821}  &         &     &       &  -19.8 &$\pm$&  19.3 &   \\
2350-0023    &  2451788.3516  & -160.3  &$\pm$&  16.6 &        &     &   & 0.74  \\
~~~~~blx/dM  &  2452523.3410  &  154.4  &$\pm$&  31.3 &        &     &   &   \\[2.9ex]
\hline
\end{tabular}

\end{center}
\end{minipage}
\begin{minipage}{\textwidth}
Notes on individual systems. 0246+0041, 0314--0111,
2241+0027, 2339--0020: variable H$\alpha$ equivalent width (EW);
0251--0000: faint, weak H$\alpha$ emission with uncertain radial
velocity measurements; 1737+5403, 2345--0014: very noisy spectrum; 
See additional notes in Table\,\ref{t-pcebfit}
\end{minipage}

\end{table*}

\begin{table*}
\caption{\label{t-rvwdms} 83 WDMS in our sample that did not show a
significant a significant radial velocity variation between their
different SDSS spectra. The complete table is available in the
electronic edition of the paper. The first column gives the SDSS
object name, the second the HJD of the spectrum, in the third column
we quote with $y$ and $n$ those spectra which are composed of
subspectra taken in different nights, the fourth and fith columns
provide the \Lines{Na}{I}{8183.27,8194.81} absorption doublet and
H$\alpha$ emission radial velocities, respectively. Blanck spaces
indicate that no radial velocity measurement could be obtained.}
\begin{tabular}{lcrclrclr}
\hline
Object  & HJD & Sub. & \multicolumn{3}{c}{RV(H$\alpha$)\,\kms} & \multicolumn{3}{c}{RV(Na)\,\kms} \\
\hline
SDSSJ001247.18+001048.7   &    2452518.4219  &  y &   0.4 &$\pm$&  26.0   &   25.6 &$\pm$&  13.8 \\  
SDSSJ001247.18+001048.7   &    2452519.3963  &  y &  34.1 &$\pm$&  46.7   &    9.1 &$\pm$&  20.0 \\
SDSSJ001726.63-002451.1   &    2452559.2853  &  y &  -4.0 &$\pm$&  16.2   &  -34.3 &$\pm$&  10.2 \\
SDSSJ001726.63-002451.2   &    2452518.4219  &  y & -34.1 &$\pm$&  11.5   &  -32.0 &$\pm$&  12.9 \\
SDSSJ001749.24-000955.3   &    2451794.7902  &  n & -47.6 &$\pm$&  13.6   &  -44.3 &$\pm$&  15.7 \\
SDSSJ001749.24-000955.3   &    2452518.4219  &  y &  -7.3 &$\pm$&  14.9   &   -8.7 &$\pm$&  11.6 \\
SDSSJ001855.19+002134.5   &    2451816.3001  &  y &  54.2 &$\pm$&  17.6   &        &     &	 \\	
SDSSJ001855.19+002134.5   &    2451892.5884  &  n &  11.0 &$\pm$&  16.3   &        &	 &       \\
\hline
\end{tabular}
\end{table*}

\section{Stellar parameters}
\label{s-stellar}
The spectroscopic data provided by the SDSS project are of sufficient
quality to estimate the stellar parameters of the WDMS presented in
this paper. For this purpose, we have developed a procedure which
decomposes the WDMS spectrum into its white dwarf and main sequence
star components, determines the spectral type of the companion by
means of template fitting, and derives the white dwarf effective
temperature (\Teff) and surface gravity ($\log g$) from spectral model
fitting. Assuming an empirical spectral type-radius relation for the
secondary star and a mass-radius relation for the white dwarf, two
independent distance estimates are calculated from the flux scaling
factors of the template/model spectra.

In the following sections, we describe in more detail the spectral
templates and models used in the decomposition and fitting, the
method adopted to fit the white dwarf spectrum, our empirical spectral
type-radius relation for the secondary stars, and the distance
estimates derived from the fits.

\begin{figure*}
\label{f-compositefit}
\includegraphics[angle=-90,width=0.49\textwidth]{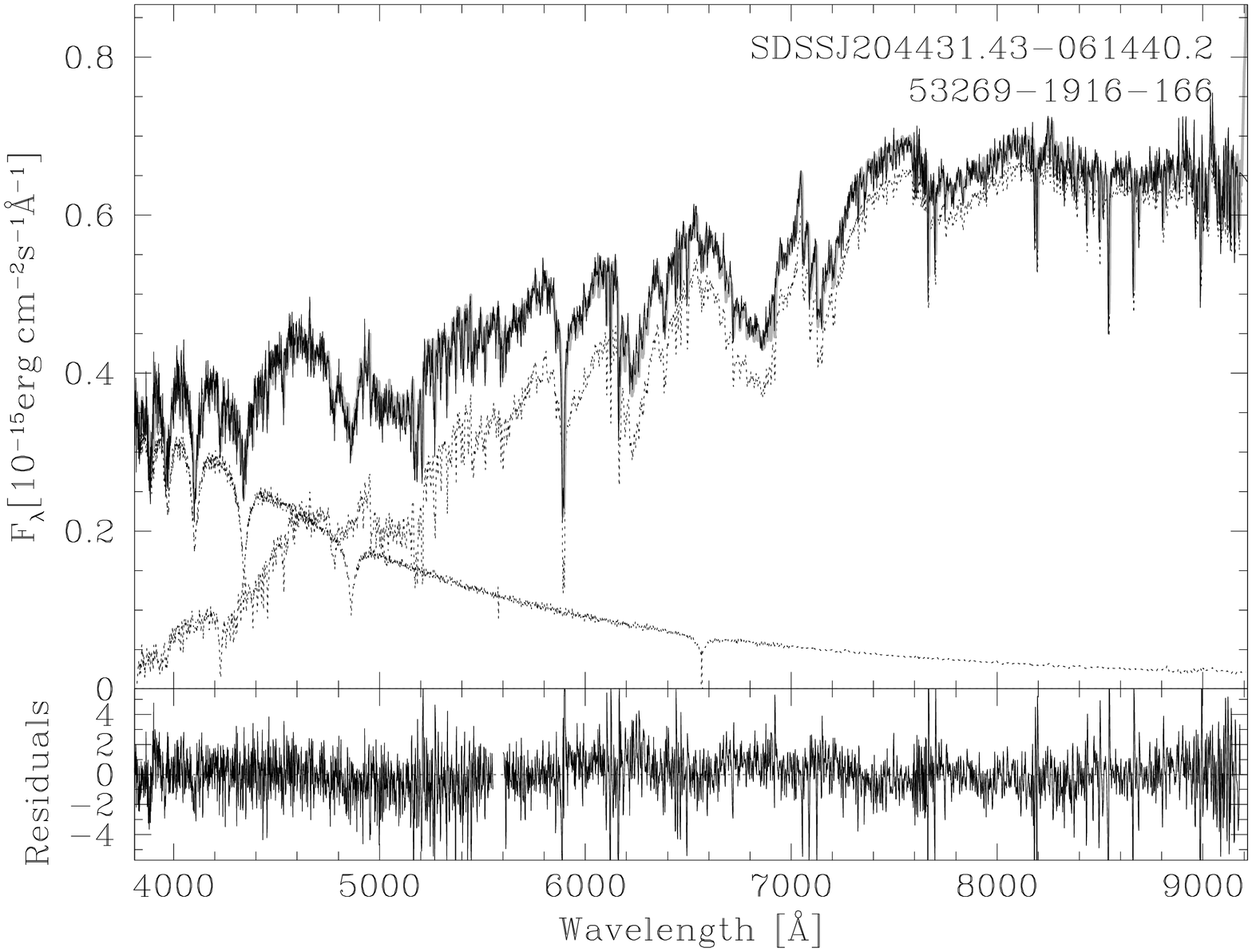}
\hfill
\includegraphics[angle=-90,width=0.49\textwidth]{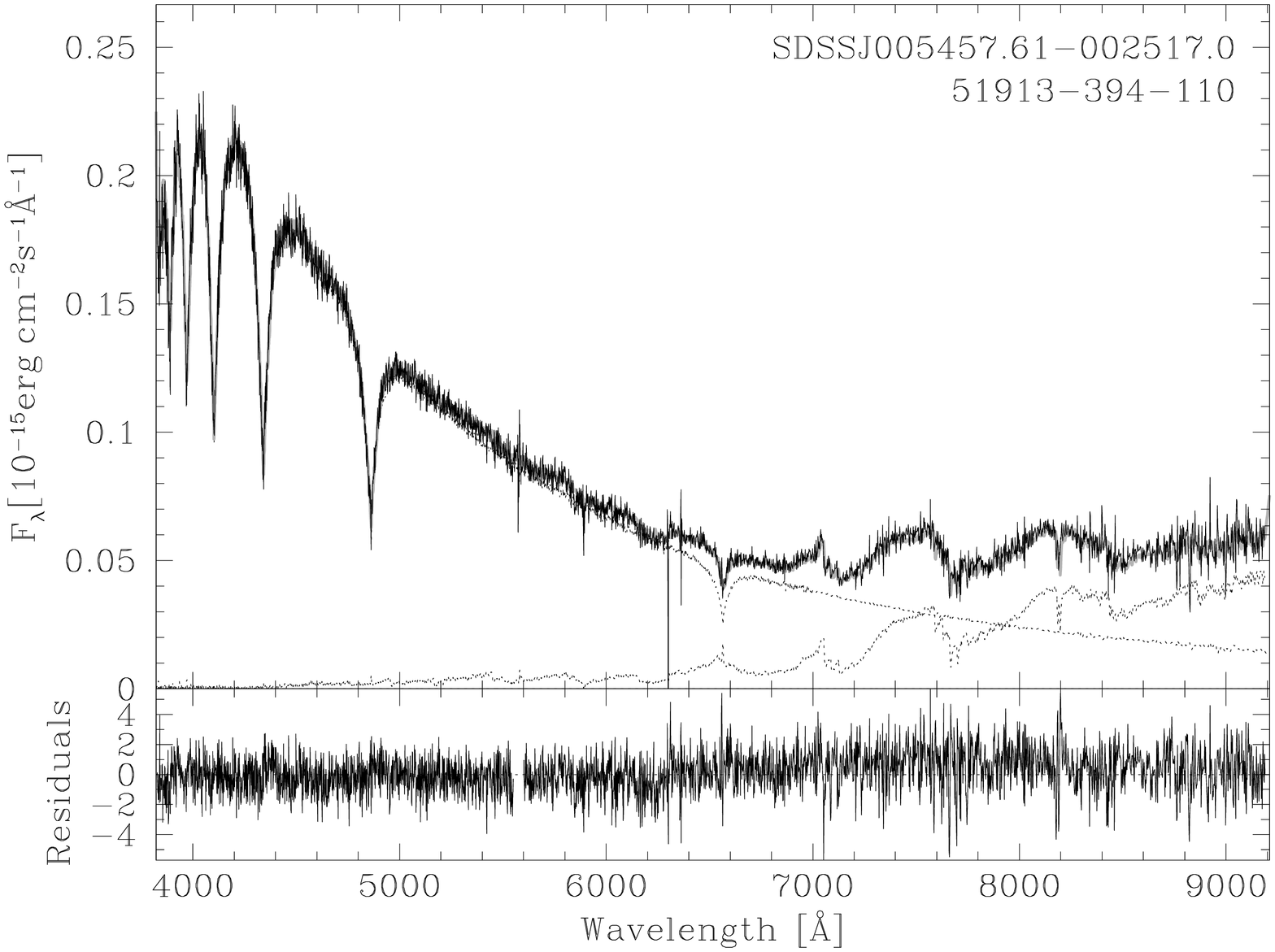}
\caption{Two-component fits to the SDSS WDMS spectra. Shown are examples
for objects with either the M-dwarf or the white dwarf dominating the
SDSS spectrum. The top panels show the WDMS spectrum as black line,
and the two templates, white dwarf and M-dwarf, as dotted lines. The
bottom panels show the residuals from the fit. The SDSS spectrum
identifies MJD, PLT and FIB are given in the plots below the object names.}

\label{f-wdfit}
\includegraphics[width=0.49\textwidth]{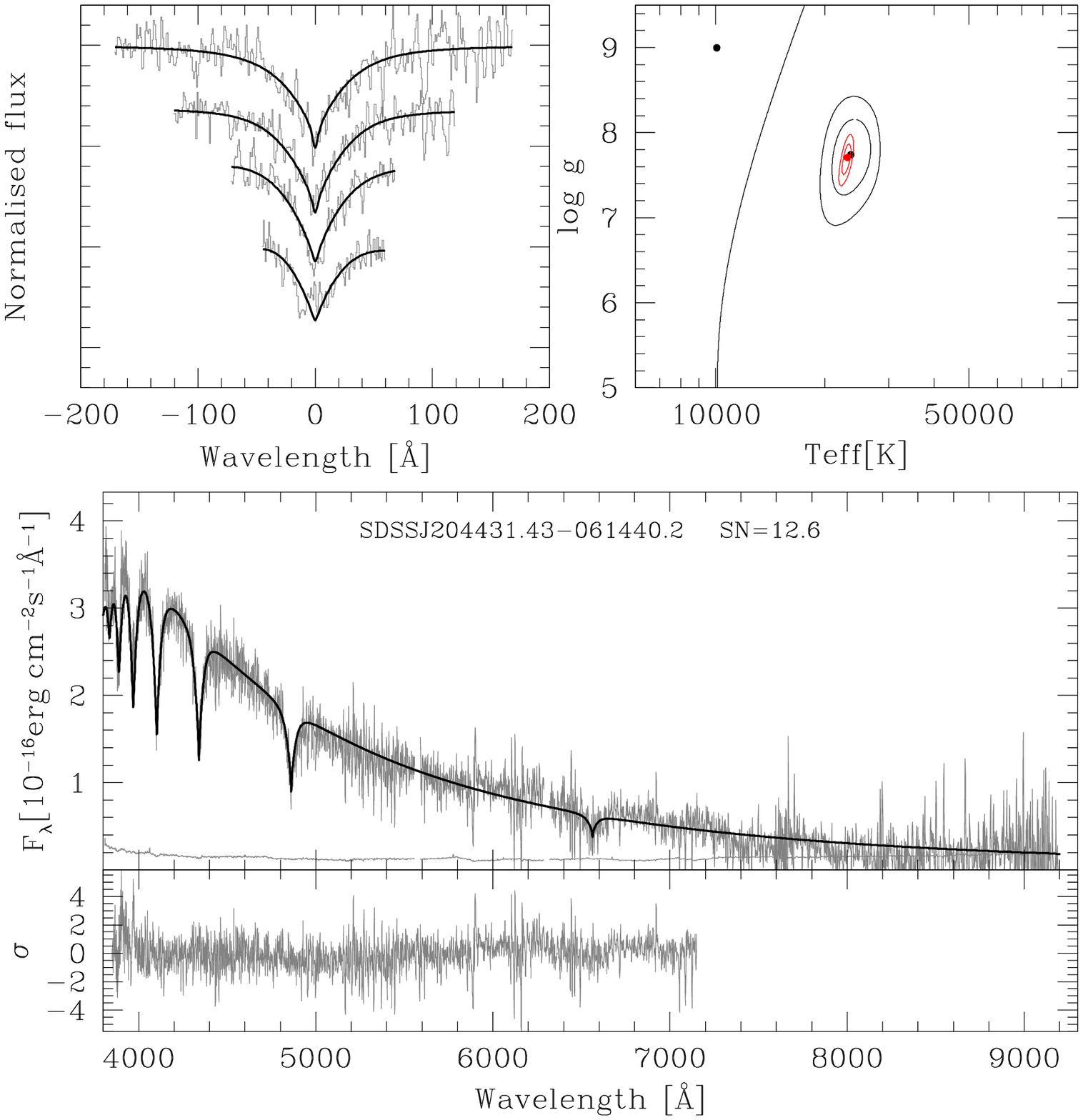}
\hfill
\includegraphics[width=0.49\textwidth]{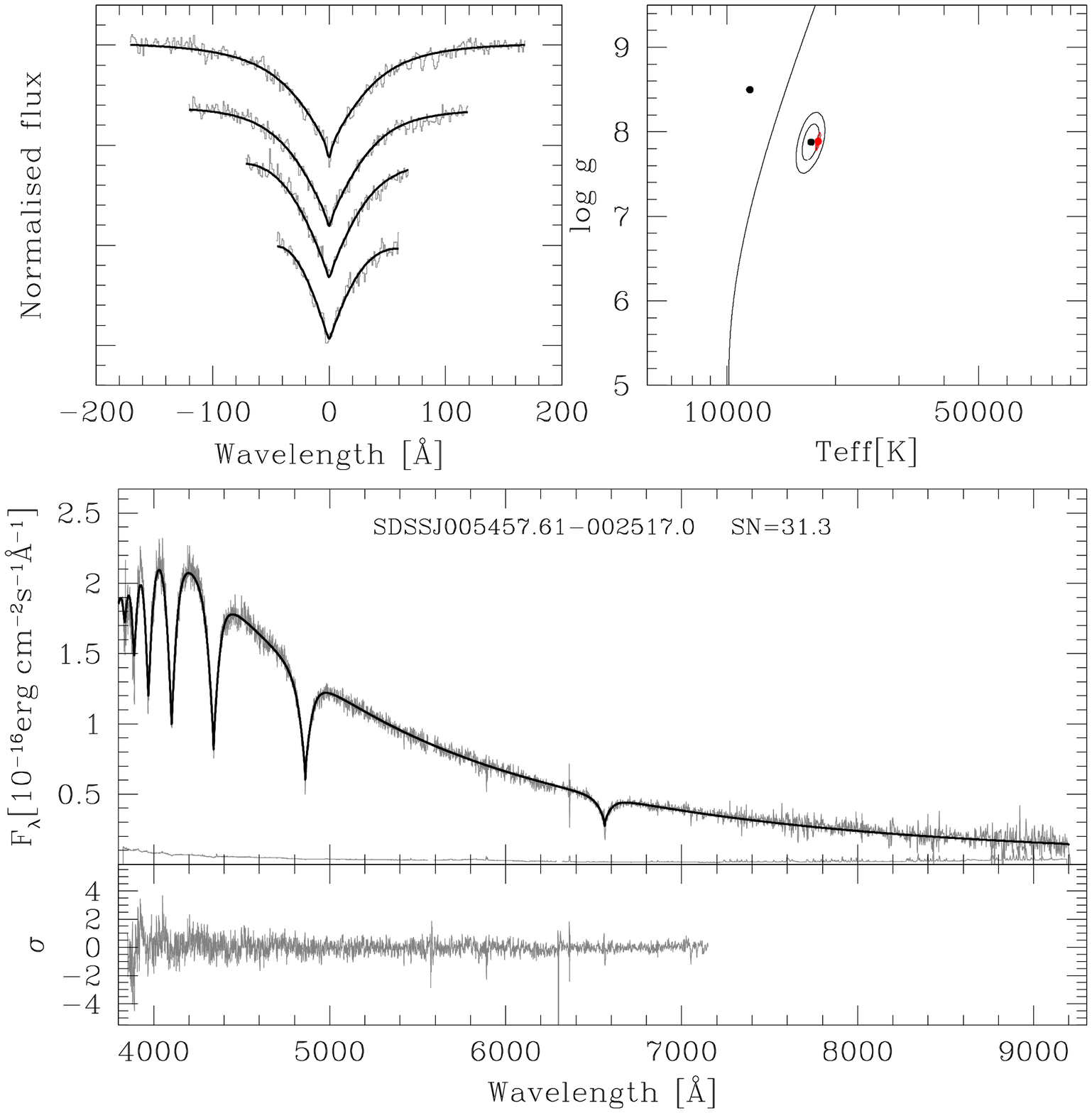}
\caption{Spectral model fits to the white dwarf components of the two
  WDMS shown in Fig.5, obtained after subtracting the best-fit M-dwarf
  template. Top left panels: best-fit (black lines) to the normalised
  H$\beta$ to H$\epsilon$ (gray lines, top to bottom) line
  profiles. Top right panels: 3, 5, and 10$\sigma$ $\chi^2$ contour
  plots in the $\Teff-\log g$ plane. The black contours refer to the
  best line profile fit, the red contours to the fit of the whole
  spectrum. The dashed line indicates the occurrence of maximum
  H$\beta$ equivalent width. The best ``hot'' and ``cold'' line
  profile solutions are indicated by black dots, the best fit to the
  whole spectrum is indicated by a red dot. Bottom panels: the
  residual white dwarf spectra resulting from the spectral
  decomposition and their flux errors (gray lines) along with the
  best-fit white dwarf model (black line) to the 3850--7150\,\AA\
wavelength
  range (top) and the residuals of the fit (gray line, bottom).  The
$\Teff$ and
  $\log g$ values listed in Table\,\ref{t-pcebfit} are determined from
  the best line profile fit. The fit to the whole spectrum is only
  used to select between the ``hot'' and ``cold'' line fit.}
\end{figure*}

\subsection{Spectral templates and models}
\label{s-templates}
In the course of decomposing/fitting the WDMS observations, we make
use of a grid of observed M-dwarf templates, a grid of observed white
dwarf templates, and a grid of white dwarf model spectra.  High
S/N ratio M-dwarf templates matching the spectral
coverage and resolution of the WDMS data were produced from a few
hundred late-type SDSS spectra from DR4. These spectra were classified
using the M-dwarf templates of \citet{beuermannetal98-1}. We averaged
the $10-20$ best exposed spectra per spectral subtype. Finally, the
spectra were scaled in flux to match the surface brightness at
7500\,\AA\ and in the TiO absorption band near 7165\,\AA, as defined
by \citet{beuermann06-1}. Recently, \citet{bochanskietal07-1}
published a library of late type stellar templates. A comparison
between the two sets of M-dwarf templates did not reveal any
significant difference.  We also compiled a library of 490 high S/N DA
white dwarf spectra from DR4 covering the entire observed range of
\Teff\ and $\log g$. As white dwarfs are blue objects, their spectra
suffer more from residual sky lines in the $I$-band. We have smoothed
the white dwarf templates at wavelengths $>7000$\,\AA\ with a
five-point box car to minimise the amount of noise added by the
residual sky lines.  Finally, we computed a grid of synthetic DA white
dwarf spectra using the model atmosphere code described by
\citet{koesteretal05-1}, covering $\log g=5.0-9.5$ in steps of 0.25 and
$\Teff=6000-100000$\,K in 37 steps nearly equidistant in
$\log(\Teff)$.

\subsection{Spectral decomposition and typing of the secondary star}
\label{s-decomposition}
Our approach is a two-step procedure. In a first step, we fitted the
WDMS spectra with a two-component model and determined the spectral
type of the M-dwarf. Subsequently, we subtracted the best-fit M-dwarf,
and fitted the residual white dwarf spectrum (Sect.\,\ref{s-wd}).  We
used an evolution strategy \citep{rechenberg94-1} to decompose the
WDMS spectra into their two individual stellar components. In brief,
this method optimises a fitness function, in this case a weighted
$\chi^2$, and allows an easy implementation of additional
constraints. Initially, we used the white dwarf model spectra and the
M-dwarf templates as spectral grids. However, it turned out that the
flux calibration of the SDSS spectra is least reliable near the blue
end of the spectra, and correspondingly, in a number of cases the
$\chi^2$ of the two-component fit was dominated by the poor match of
the white dwarf model to the observed data at short wavelengths. As we
are in this first step not yet interested in the detailed parameters of
the white dwarf, but want to achieve the best possible fit of the
M-dwarf, we decided to replace the white dwarf models by observed
white dwarf templates. The large set of observed white dwarf
templates, which are subject to the same observational issues as the
WDMS spectra, provided in practically all cases a better match in the
blue part of the WDMS spectrum. From the converged white dwarf plus dM
template fit to each WDMS spectrum (see Fig.\,5), we recorded the
spectral type of the secondary star, as well as the flux scaling
factor between the M-star template and the observed spectrum. The
typical uncertainty in the spectral type of the secondary star is
$\pm0.5$ spectral class. The spectral types determined from the
composite fits to each individual spectrum are listed in
Table\,\ref{t-pcebfit} for the PCEBs in the analysed sample, and in
the electronic edition of this paper for the remaining
WDMS (see Table\,\ref{t-wdmsfit}). Inspection of those tables shows that for the vast majority of
systems, the fits to the individual spectra give consistent
parameters. We restricted the white dwarf fits to  WDMS containing a DA
primary, consequently no white dwarf parameters are  provided for those WDMS
containing DB or DC white dwarfs.

\subsection{White dwarf parameters}
\label{s-wd}
Once the best-fit M-dwarf template has been determined and scaled
appropriately in flux, it is subtracted from the WDMS spectrum. The
residual white dwarf spectrum is then fitted with the grid of DA
models described in Sect.\,\ref{s-templates}. Because of the
uncertainties in the flux calibration of the SDSS spectra and the flux
residuals from the M-star subtraction, we decided to fit the
normalised H$\beta$ to H$\epsilon$ lines and omitted H$\alpha$ where
the residual contamination from the secondary star was largest. While
the sensitivity to the surface gravity increases for the higher Balmer
lines \citep[e.g.][]{kepleretal06-1}, we decided not to include them
in the fit because of the deteriorating S/N and the unreliable flux
calibration at the blue end. We determined the best-fit
$\Teff$ and $\log g$ from a bicubic spline interpolation to the
$\chi^2$ values on the $\Teff-\log g$ grid defined by our set of model
spectra. The associated 1$\sigma$ errors were determined from
projecting the contour at $\Delta\chi^2=1$ with respect to the
$\chi^2$ of the best fit onto the $\Teff$ and $\log g$ axes and
averaging the resulting parameter range into a symmetric error bar.

The equivalent widths of the Balmer lines go through a maximum near
$\Teff=13\,000$\,K, with the exact value being a function of $\log
g$. Therefore, $\Teff$ and $\log g$ determined from Balmer line
profile fits are subject to an ambiguity, often referred to as ``hot''
and ``cold'' solutions, i.e. fits of similar quality can be achieved
on either side of the temperature at which the maximum equivalent
width is occurring. We measured the H$\beta$ equivalent width in all
the model spectra within our grid, and fitted the dependence of the
temperature at which the maximum equivalent width of H$\beta$ occurs
by a second-order polynomial, 

\begin{equation}
\Teff(\mathrm{EW[H\beta]_{max}})=20361-3997\log g+390(\log g)^2
\label{e-maxew}
\end{equation}

Parallel to the fits to the normalised line profiles, we fit the grid
of model spectra to the white dwarf spectrum over the wavelength range
$3850-7150$\,\AA\ (see Fig.~\ref{f-wdfit}). The red end of the SDSS
spectra, where the distortion from the M-dwarf subtraction is strongest
is excluded from the fit. We then use the $\Teff$ and $\log g$ from
the fits to the whole spectrum, continuum plus lines, to select the
``hot'' or ``cold'' solution from the line profile fits. In the
majority of cases, the solution preferred by the fit to the whole
spectrum has a substantially lower $\chi^2$ than the other solution,
corroborating that it is likely to be the physically correct
choice. In a few cases, the best-fit $\Teff$ and $\log g$ from the
whole spectrum are close to the maximum equivalent width given by
Eq.\ref{e-maxew}, so that the choice between the two line profile
solutions is less well constrained. However, in most of those cases,
the two solutions from the line profile fits overlap within their
error bars, so that the final choice of $\Teff$ and $\log g$ is not
too badly affected.

Once that $\Teff$ and $\log g$ are determined from the best line
profile fit, we use an updated version of
\citeauthor{bergeronetal95-2}'s (1995) tables to calculate the mass
and the radius of the white dwarf. Table\,\ref{t-pcebfit} reports
\Teff, $\log g$, and the white dwarf masses for the PCEBs in our
sample, while the results for the remaining WDMS can be found in the
electronic edition (Table\,\ref{t-wdmsfit}). We have carefully inspected each individual
composite fit, and each subsequent fit to the residual white dwarf
spectrum, and are confident that we have selected the correct solution
in the majority of cases. Some doubt remains primarily for a few
spectra of very low signal-to-noise ratio. 
The fact that we have analysed at least two SDSS spectra for
each system allows us to assess the robustness of our spectral
decomposition/fitting method. Inspection of Table\,\ref{t-pcebfit}
shows that the system parameter of a given system, as determined from
several different SDSS spectra, generally agree well within the quoted
errors, confirming that our error estimate is realistic.

\subsection{An empirical spectral type-radius relation for M stars}
\label{s-sp_r}

\begin{figure}
\includegraphics[width=0.49\textwidth]{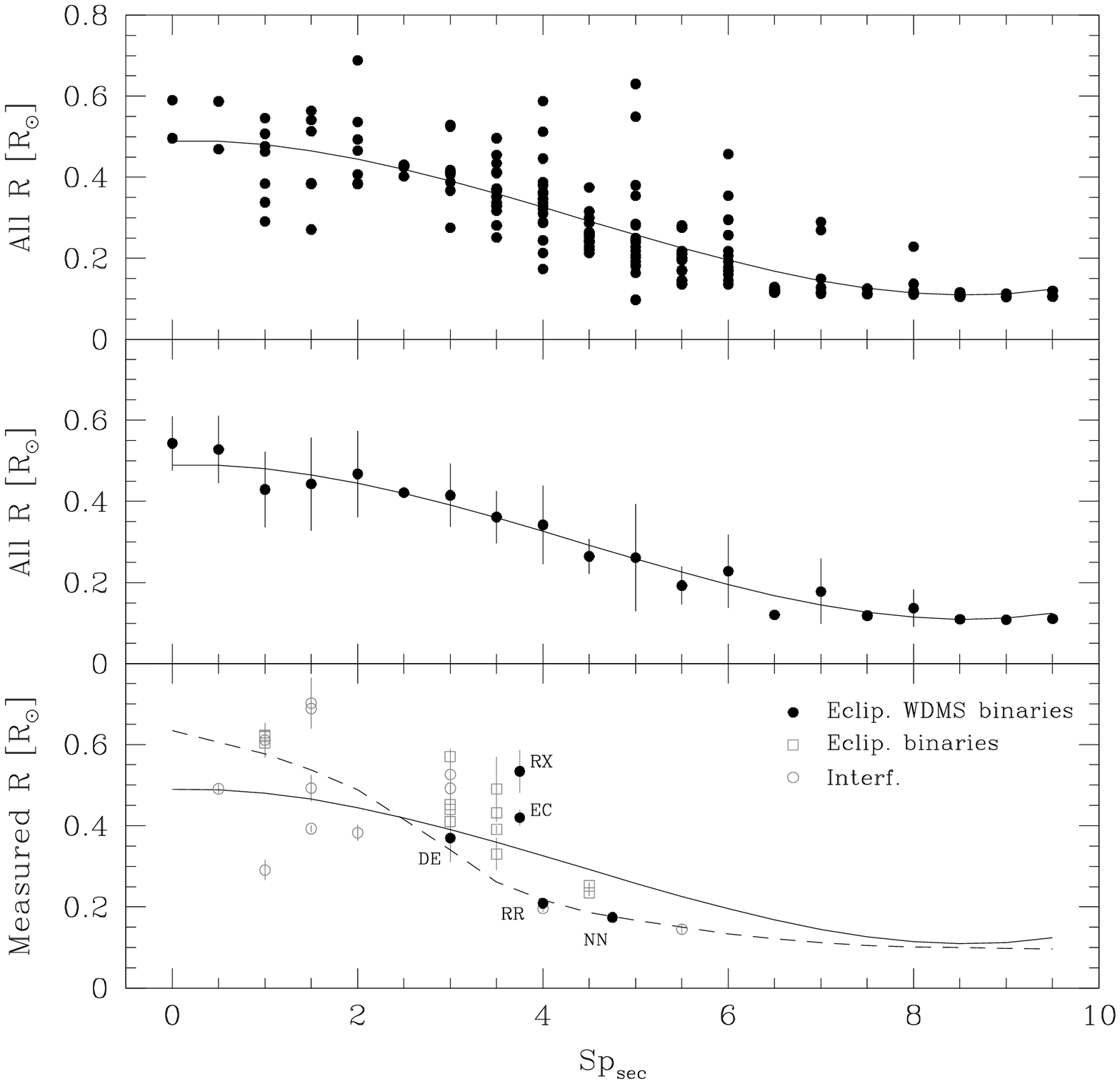}
\caption{Top panel: indirectly measured radii of M-dwarfs vs spectral
type. Our empirical $Sp-R$ relation is given by a third order polynomial
fit (solid line) to these data. Middle panel: mean radii and
corresponding standard deviations obtained by averaging the radii
in the top panel for each spectral type. Our $Sp-R$ relation is again
superimposed.  Bottom panel: directly measured radii of M-dwarfs,
again our empirical $Sp-R$ relation, the dashed line is the theoretical
$Sp-R$ relation from \citet{baraffeetal98-1}. M-dwarf radii from the
eclipsing WDMS  RR\,Cae, NN\,Ser, DE\,CVn, RX\,J2130.6+4710
and EC\,13471--1258 are shown as solid dots.}
\label{f-SpR}
\end{figure}

\begin{table}
    \centering
    \begin{tabular}{|c|c|c|c|c|c|}
    \hline
    \hline
    Sp & R$_\mathrm{mean}$ (R$_\odot$) &  R$_\mathrm{\sigma}$ (R$_\odot$)  & R$_\mathrm{fit}$ (R$_\odot$) & M$_\mathrm{fit}$ (M$_\odot$) & $\Teff$ (K) \\
    \hline
    \\
   M0.0 & 0.543 & 0.066 &  0.490 & 0.472 &    3843\\
   M0.5 & 0.528 & 0.083 &  0.488 & 0.471 &    3761\\
   M1.0 & 0.429 & 0.094 &  0.480 & 0.464 &    3678\\
   M1.5 & 0.443 & 0.115 &  0.465 & 0.450 &    3596\\
   M2.0 & 0.468 & 0.106 &  0.445 & 0.431 &    3514\\
   M2.5 & 0.422 & 0.013 &  0.420 & 0.407 &    3432\\
   M3.0 & 0.415 & 0.077 &  0.391 & 0.380 &    3349\\
   M3.5 & 0.361 & 0.065 &  0.359 & 0.350 &    3267\\
   M4.0 & 0.342 & 0.096 &  0.326 & 0.319 &    3185\\
   M4.5 & 0.265 & 0.043 &  0.292 & 0.287 &    3103\\
   M5.0 & 0.261 & 0.132 &  0.258 & 0.255 &    3020\\
   M5.5 & 0.193 & 0.046 &  0.226 & 0.225 &    2938\\
   M6.0 & 0.228 & 0.090 &  0.195 & 0.196 &    2856\\
   M6.5 & 0.120 & 0.005 &  0.168 & 0.170 &    2773\\
   M7.0 & 0.178 & 0.080 &  0.145 & 0.149 &    2691\\
   M7.5 & 0.118 & 0.009 &  0.126 & 0.132 &    2609\\
   M8.0 & 0.137 & 0.046 &  0.114 & 0.120 &    2527\\
   M8.5 & 0.110 & 0.004 &  0.109 & 0.116 &    2444\\
   M9.0 & 0.108 & 0.004 &  0.112 & 0.118 &    2362\\
   M9.5 & 0.111 & 0.008 &  0.124 & 0.130 &    2281\\
    \hline 
    \end{tabular}
\caption{Empirical $Sp-R$, $Sp-M$ and $Sp-\Teff$ relations (R$_\mathrm{fit}$,
M$_\mathrm{fit}$, \Teff) found in this work. R$_\mathrm{mean}$ and
R$_\mathrm{\sigma}$ represent the mean radii and their standard
deviation obtained from the sample of M-dwarfs described in
Sect.\,\ref{s-sp_r}.}
\label{t-Sp-R-M}
\end{table}

In order to use the flux scaling factor between the observed WDMS
spectra and the best-fit M-dwarf templates for an estimate of the
distance to the system (Sect.\,\ref{s-distances}), it is necessary to
assume a radius for the secondary star. Since we have  determined
the spectral types of the companion stars from the SDSS spectra 
(Sect.\,\ref{s-decomposition}), we require 
a spectral type-radius relation ($Sp-R$) for M-dwarfs. The community 
working on cataclysmic variables has previously had interest
in such a relation \citep[e.g.][]{mateoetal85-1,
caillault+patterson90-1}, but while \citet{baraffe+chabrier96-1}
derived theoretical mass/radius/effective temperature-spectral type
relationships for single M-dwarfs, relatively little observational
work along these lines has been carried out for field low mass
stars. In contrast to this, the number of low mass stars with accurate
mass and radius measurements has significantly increased over the past few
years \citep[see e.g. the review by][]{ribas06-1}, and it appears that
for masses below the fully convective boundary stars follow  the 
theoretical models by \citet{baraffeetal98-1} relatively well. However, for
masses $\ga0.3$\,M$_\odot$, observed radii exceed the predicted ones. 
Stellar activity \citep[e.g.][]{lopez-morales07-1} or
metallicity effects \citep[e.g.][]{bergeretal06-1} were identified as
possible causes.

\begin{table*}
\setlength{\tabcolsep}{1.0ex}
\caption{\label{t-wdmsfit}Stellar parameters of the remaining WDMS
identified in our sample, as determined from spectral modelling. The
complete table can be found in the electronic edition. Given are, from
left to right: SDSS object name, MJD, fiber and plate number of the
spectrum, white dwarf effective temperature and error, white dwarf
surface gravity and error, mass of the white dwarf and error, distance
to the white dwarf and error, spectral type of the secondary star,
distance to the secondary and error, flag (we refer by and $s$ and $e$
those systems which have been studied previously by
\citet{silvestrietal06-1} and \citet{eisensteinetal06-1}, by $re$
those systems whose binary components are resolved), and notes.}

\begin{tabular}{lcrclrclrclrclrcl}
\hline
Object    &	MJD  &	plate & fiber	&   T(k) &  err	&  $\log g$ & err & M(M$_{\bigodot}$) & err & d$_\mathrm{wd}$(pc) & err & Sp & d$_\mathrm{sec}$(pc) & err & flag & notes \\
\hline
 SDSSJ000442.00-002011.6 & 51791  &   387  &     24 &        - &     - &      - &     - &     - &     - &     - &     -  &   0 &   2187 &   236 &   re	  & 1	  \\				
 			 & 52943  &   1539 &     21 &        - &     - &      - &     - &     - &     - &     - &     -  &   0 &   2330 &   251 &	  &	  \\				
 SDSSJ001029.87+003126.2 & 51793  &   388  &    545 &        - &     - &      - &     - &     - &     - &     - &     -  &   2 &   1639 &   339 &   s,re  &  	  \\				
 			 & 52518  &   687  &    347 &    13904 &  3751 &   8.43 &  1.16 &  0.88 &  0.62 &   781 &   605  &   2 &   1521 &   314 &         &	  \\				
 SDSSJ001247.18+001048.7 & 52518  &   687  &    395 &    18542 &  5645 &   8.75 &  0.80 &  1.07 &  0.41 &   661 &   446  &   3 &    830 &   132 &   e	  &  	  \\				
 			 & 52519  &   686  &    624 &    32972 &  7780 &   8.61 &  1.11 &  1.01 &  0.54 &  1098 &   930  &   3 &    936 &   149 &	  &	  \\				
 SDSSJ001749.24-000955.3 & 51795  &   389  &    112 &    72136 &  3577 &   8.07 &  0.14 &  0.77 &  0.07 &   532 &    60  &   2 &    684 &   142 &   s,e   &	  \\
 			 & 52518  &   687  &    109 &    69687 &  4340 &   7.61 &  0.20 &  0.59 &  0.07 &   784 &   127  &   2 &    659 &   136 &	  &	  \\
 SDSSJ001726.63-002451.1 & 52559  &   1118 &    280 &    12828 &  2564 &   8.00 &  0.46 &  0.61 &  0.29 &   422 &   120  &   4 &    579 &   172 &   s,e	  & 	  \\				
 			 & 52518  &   687  &    153 &    13588 &  1767 &   8.11 &  0.38 &  0.68 &  0.24 &   424 &   106  &   4 &    522 &   155 &	  &	  \\				
 SDSSJ001855.19+002134.5 & 51816  &   390  &    385 &        - &     - &      - &     - &     - &     - &     - &     -  &   3 &   1186 &   189 &         &  	  \\				
 			 & 51900  &   390  &    381 &    14899 &  9266 &   9.12 &  1.03 &  1.26 &  0.54 &   445 &   330  &   3 &   1249 &   199 &	  &	  \\				
 			 & 52518  &   687  &    556 &    10918 &  4895 &   8.64 &  2.01 &  1.00 &  1.06 &   539 &   247  &   3 &   1087 &   173 &	  &	  \\
\hline

\multicolumn{17}{p{\textwidth}}{(1) Possible K secondary star}
\end{tabular}
\end{table*}


Besides the lack of extensive observational work on the $Sp-R$
relation of single
M-dwarfs, our need for an M-dwarf $Sp-R$ relation in the context of WDMS
faces a number of additional problems. A fraction of the WDMS in our
sample have undergone a CE phase, and are now
short-period binaries, in which the secondary star is tidally locked
and hence rapidly rotating. This rapid rotation will enhance the stellar
activity in a similar fashion to the short-period eclipsing M-dwarf
binaries used in the $M-R$ relation work mentioned above. In addition,
it is difficult to assess the age\footnote{\label{n-age} In principle,
an age estimate can be derived by adding the white dwarf cooling age
to the main sequence life time of the white dwarf progenitor. This
involves the use of an initial mass-final mass relation for the white
dwarf, e.g. \citet{dobbieetal06-1}, which will not be strictly valid
for those WDMS that underwent a CE evolution. Broadly
judging from the distribution of white dwarf temperatures and masses
in Fig.\,\ref{f-histo}, most WDMS in our sample should be older than
1\,Gyr, but the data at hand does not warrant a more detailed
analysis.} and metallicity of the secondary stars in our WDMS sample.

With the uncertainties on stellar parameters of single M-dwarfs and
the potential additional complications in WDMS in mind, we decided to
derive an ``average'' $Sp-R$ relation for M-dwarfs irrespective of their
ages, metallicities, and activity levels. The primary purpose of this
is to provide distance estimates based on the flux scaling
factors in Eq.\,\ref{e-secdist}, but also to assess potential
systematic peculiarities of the secondary stars in the WDMS. 

We have compiled spectral types and radii of field M-dwarfs from
\citet{berriman+reid87-1}, \citet{caillault+patterson90-1}, 
\citet{leggettetal96-1}, \citet{delfosseetal99-1}, \citet{letoetal00-1},
\citet{laneetal01-1}, \citet{segransanetal03-1}, 
\citet{maceroni+montalban04-1}, \citet{creeveyetal05-1}, \citet{pontetal05-1},
\citet{ribas06-1}, \citet{bergeretal06-1}, \citet{bayless+orosz06-1} 
and \citet{beattyetal07-1}. These
data were separated into two groups, namely stars with directly
measured radii (in eclipsing binaries or via interferometry) and stars
with indirect radii determinations (e.g. spectrophotometric). We complemented this
sample with spectral types, masses, effective temperatures and
luminosities from \citet{delfosseetal98-1}, \citet{leggettetal01-1},
\citet{berger02-1}, \citet{golimowskietal04-1}, \citet{cushingetal05-1}
and \citet{montagnieretal06-1}, calculating radii from $ L = 4\pi R^{2}
\sigma \Teff^{4} $ and/or \citeauthor{caillault+patterson90-1}'s
(\citeyear{caillault+patterson90-1}) mass-luminosity and mass-radius
relations.

Figure\,\ref{f-SpR} shows our compilation of indirectly determined
radii as a function of spectral type (top panel) as well as those from
direct measurements (bottom panel). A large scatter in radii is
observed at all spectral types except for the very late M-dwarfs, where
only few measurements are available. It is interesting that
the amount of scatter is comparable for both groups of M-dwarfs, those
with directly measured radii and those with indirectly determined
radii. This underlines that systematic effects intrinsic to the stars cause
a large spread in the $Sp-R$ relation even for the objects with accurate
measurements. In what follows, we use the indirectly measured radii as
our primary sample, as it contains a larger number of stars and
extends to later spectral types. The set of directly measured radii
are used as a comparison to illustrate the $Sp-R$ distribution of stars
where the systematic errors in the determination of their radii is
thought to be small. We determine an $Sp-R$ relation from fitting the
indirectly determined radius data with third order polynomial,  

\begin{equation}
R = 0.48926 -~ 0.00683~\mathrm{Sp} -~ 0.01709~\mathrm{Sp}^{2} +~ 0.00130~\mathrm{Sp}^{3} 
\label{e-sp-r}
\end{equation}

\noindent
The spectral type is not a physical quantity, and strictly speaking,
this relation is only defined on the existing spectral classes.
This fit agrees well with the average of the radii in each spectral
class (Fig.\,\ref{f-SpR}, middle panel, where the errors are the
standard deviation from the mean value). The radii from the polynomial
fit are reported in Table\,\ref{t-Sp-R-M}, along with the average
radii per spectral class. Both the radii from the polynomial fit and
the average radii show a marginal upturn at the very latest spectral
types, which should not be taken too seriously given the small number of
data involved. 

We compare in Fig.\,\ref{f-SpR} (bottom panel) the directly measured
radii with our $Sp-R$ relation. It is apparent that also stars with
well-determined radii show a substantial amount of scatter, and are
broadly consistent with the empirical $Sp-R$ relation determined from the
indirectly measured radii. As a test, we included the directly
measured radii in the fit described above, and did not find any
significant change compared to the indirectly measured radii alone.

For a final assessment on our empirical $Sp-R$ relation, specifically in
the context of WDMS, we have compiled from the literature the radii of
M-dwarfs in the eclipsing WDMS RR\,Cae \citep{maxtedetal07-1}, NN\,Ser
\citep{haefneretal04-1}, DE\,CVn \citep{vandenbesselaaretal07-1},
RX\,J2130.6+4710 \citep{maxtedetal04-1}, and EC\,13471--1258
\citep{odonoghueetal03-1}, (Fig.\,~\ref{f-SpR}, bottom panel). Just as
the accurate radii determined from interferometric observations of
M-dwarfs or from light curve analyses of eclipsing M-dwarf binaries,
the radii of the secondary stars in WDMS display a substantial  amount
of scatter.

\subsubsection{Comparison with the theoretical Sp-R relation from
  \citet{baraffeetal98-1}} We compare in the bottom panel of
Fig.\,\ref{f-SpR} our empirical $Sp-R$ relation with the theoretical
prediction from the evolutionary sequences of \citet{baraffeetal98-1},
where the spectral type is based on the $I-K$ colour of the
\texttt{PHOENIX} stellar atmosphere models coupled to the stellar
structure calculations. The theoretical $Sp-R$ relation displays
substantially more curvature than our empirical relation, predicting
larger radii for spectral types $\la$M2, and significantly smaller
radii in the range M3--M6. The two relations converge at late spectral
types (again, the upturn in the empirical relation for $>$M8.5 should
be ignored as an artifact from our polynomial fit). The ``kink'' in
the theoretical relation seen around M2 is thought to be a consequence
of H$_2$ molecular dissociation \citep{baraffe+chabrier96-1}. The
large scatter of the directly determined radii of field M-dwarfs as
well of M-dwarfs in eclipsing WDMS could be related to two types of
problem, that may have a common underlying cause. (1) In eclipsing
binaries, the stars are forced to extremely rapid rotation, which is
thought to increase stellar activity that is likely to affect the
stellar structure, generally thought to lead to an increase in radius
\citep{spruit+weiss86-1, mullan+macdonald01-1, chabrieretal07-1}, and
(2) the spectral types in our compilations of radii are determined
from optical spectroscopy, and may differ to some extent from the
spectral type definition based on $I-K$ colours as used in the
\citet{baraffeetal98-1} models. Furthermore, stellar activity is
thought to affect not only the radii of the stars, but also their
luminosity, surface temperatures, and hence spectral types. The effect
of stellar activity is discussed in more detail in
Sect.\,\ref{s-distdist}.

\subsubsection{$Sp-\Teff$ and $Sp-M$ relations}
For completeness, we fitted the spectral type-mass data and
the spectral type-effective temperature data compiled from the
literature listed above, and fitted the $Sp-M$ and $Sp-\Teff$
relations with a third-order polynomial and a first-order polynomial,
respectively. The results from the fits are reported in 
Table\,\ref{t-Sp-R-M}, and will be used in this paper only for
estimating upper limits to the orbital periods of our PCEBs
(Sect.\,\ref{s-porb}) and when discussing the possibility of stellar
activity on the WDMS secondary stars in Sect.\,\ref{s-distdist}.

\subsection{Distances}
\label{s-distances}
The distances to the WDMS can be estimated from the best-fit flux scaling
factors of the two spectral components. For the white dwarf, 

\begin{equation}
\frac{f_\mathrm{wd}}{F_\mathrm{wd}} = \pi \left (\frac{R_\mathrm{wd}}{d_\mathrm{wd}}
\right)^{2} 
\label{e-wddist}
\end{equation}

\noindent
where $f_\mathrm{wd}$ is the observed flux of the white dwarf,
$F_\mathrm{wd}$ the astrophysical flux at the stellar surface as given
by the model spectra, $R_\mathrm{wd}$ is the white dwarf radius and
$d_\mathrm{wd}$ is the distance to the WD. For the secondary star,

\begin{equation}
\frac{f_\mathrm{sec}}{F_\mathrm{sec}} = \left (\frac{R_\mathrm{sec}}{d_\mathrm{sec}}
\right)^{2}  
\label{e-secdist}
\end{equation}

where $f_\mathrm{sec}$ is the observed M-dwarf flux, $F_\mathrm{sec}$
the flux at the stellar surface, and $R_\mathrm{sec}$ and
d$_\mathrm{sec}$ are the radius and the distance to the secondary
respectively.

The white dwarf radii are calculated from the best-fit $\Teff$ and
$\log g$ as detailed in Sect.\,\ref{s-wd}. The secondary star radii
are taken from Table\,\ref{t-Sp-R-M} for the best-fit spectral type. The
uncertainties of the distances are based on the errors in
$R_\mathrm{wd}$, which depend primarily on the error in $\log g$, and
in $R_\mathrm{sec}$, where we assumed the standard deviation from
Table\,\ref{t-Sp-R-M} for the given spectral type. 
Table\,\ref{t-pcebfit} lists the values $d_\mathrm{wd}$ and 
$d_\mathrm{sec}$ obtained for our PCEBs. The remaining 112 WDMS's distances
can be found in the electronic edition (Table\,\ref{t-wdmsfit}).

\section{Discussion}
\label{s-discussion}

\begin{figure*}
\includegraphics[angle=-90,width=\textwidth]{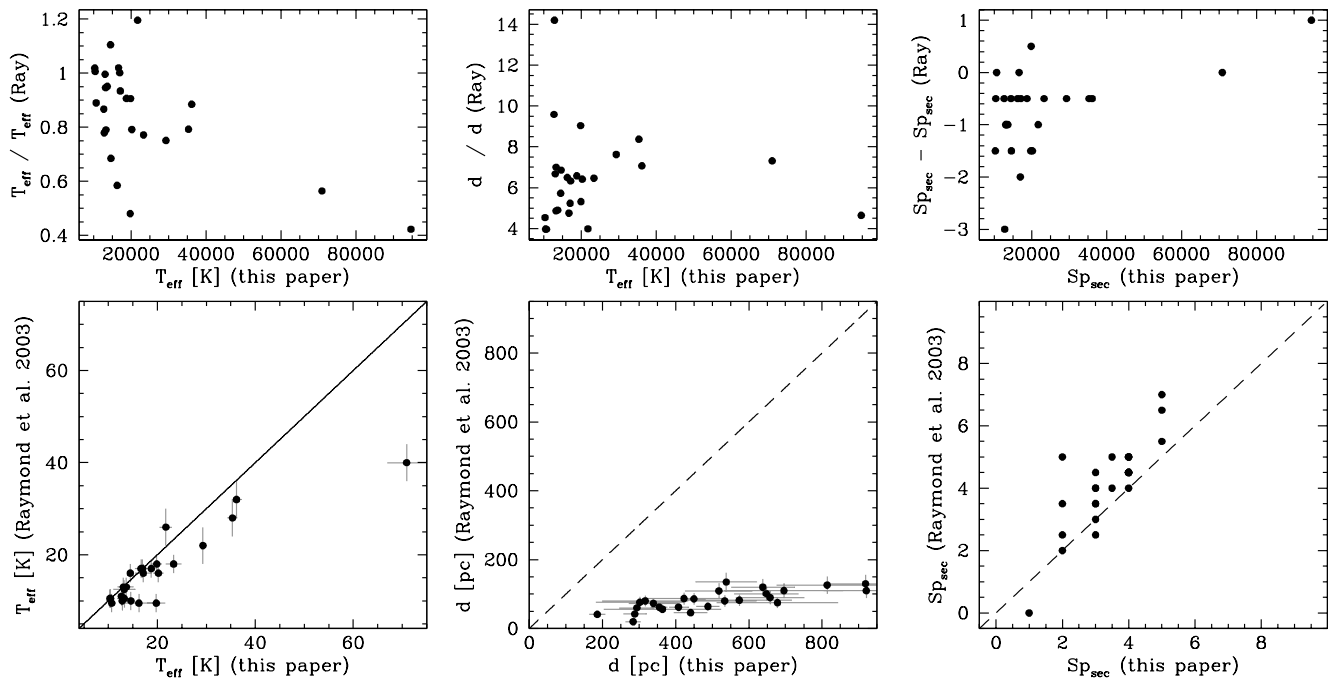}
\caption{\label{f-comp_ray}Comparison of the white dwarf effective
temperatures, distances based on the white dwarf fit, and the spectral
types of the secondary stars determined from our fits
(Sect.\,\ref{s-decomposition}, \ref{s-wd} and Table\,\ref{t-pcebfit}),
and those of \citet{raymondetal03-1}. Top panels, from left to right:
the ratio in \Teff, the ratio in $d$, and the difference in the
secondary's spectral types from the two studies as a function of the
white dwarf temperature.}
\end{figure*}

\subsection{H$\alpha$ vs \Ion{Na}{I} radial velocities}
\label{s-rv-porb}

As mentioned in Sect.\,\ref{s-PCEB}, a few systems in
Table\,\ref{t-rv} show considerable differences between their
H$\alpha$ and \Ion{Na}{I} radial velocities. More specifically, while
both lines clearly identify these systems as being radial velocity
variable, and hence PCEBs or strong PCEB candidates, the actual radial
velocities of H$\alpha$ and \Ion{Na}{I} differ for a given SDSS
spectrum by more than their errors.

In close PCEBs with short orbital periods the H$\alpha$ emission is
typically observed to arise from the hemisphere of the companion star
facing the white dwarf. Irradiation from a hot white dwarf is the most
plausible mechanism to explain the anisotropic H$\alpha$ emission,
though also a number of PCEBs containing rather cool white dwarfs are
known to exhibit concentrated H$\alpha$ emission on the inner
hemisphere of the companion stars \citep[e.g.][]{marsh+duck96-1,
maxtedetal06-1}. The anisotropy of the H$\alpha$ emission results in
its radial velocity differing from other photospheric features that
are (more) isotropically distributed over the companion stars, such as
the \Ion{Na}{I} absorption. In general, the H$\alpha$ emission line
radial velocity curve will then have a lower amplitude than that of
the \Ion{Na}{I} absorption lines, as H$\alpha$ originates closer to
the centre of mass of the binary system. In addition, the strength 
of H$\alpha$ can vary greatly due to different geometric projections 
in high inclination systems. More complications are added in the 
context of SDSS spectroscopy, where the individual spectra have typical 
exposure times of 45--60min, which will result in the smearing of the
spectral features in the short-period PCEBs due to the sampling of different 
orbital phases. This problem is exacerbated in the case that the SDSS 
spectrum is combined from exposures taken on different nights (see
Sect.\,\ref{s-PCEB}). Finally, the H$\alpha$ emission from the
companion may substantially increase during a flare, which will
further enhance the anisotropic nature of the emission.

Systems in which the H$\alpha$ and \Ion{Na}{I} radial velocities
differ by more than 2 $\sigma$ are: SDSS\,J005245.11-005337.2,
SDSS\,J024642.55+004137.2, SDSS\,J030904.82-010100.8,
SDSS\,J031404.98-011136.6, and SDSS\,J172406.14+562003.0. Of these,
SDSS\,J0246+0041, SDSS\,J0314-0111, and SDSS\,J1724+5620 show
large-amplitude radial velocity variations and substantial changes in
the equivalent width of the H$\alpha$ emission line, suggesting that
they are rather short orbital period PCEBs with moderately high
inclinations, which most likely explains the observed differences
between the observed H$\alpha$ and \Ion{Na}{I} radial
velocities. Irradiation is also certainly important in
SDSS\,J1724+5620 which contains a hot ($\simeq36000$\,K) white
dwarf. SDSS\,J0052-0053 displays only a moderate radial velocity
amplitude, and while the H$\alpha$ and \Ion{Na}{I} radial velocities
display a homogeneous pattern of variation (Fig.\,\ref{f-Na} and
\ref{f-Ha}), H$\alpha$ appears to have a larger amplitude which is not
readily explained. Similar discrepancies have been observed
e.g. in the close magnetic WDMS binary WX\,LMi, and were thought to be
related to a time-variable change in the location of the H$\alpha$
emission \citep{vogeletal07-1}. Finally, SDSS\,J0309-0101 is rather
faint ($g=20.4$), but has a strong H$\alpha$ emission that allows
reliable radial velocity measurements that identify the system as a
PCEB. The radial velocities from the \Ion{Na}{I} doublet are more
affected by noise, which probably explains the observed radial
velocity discrepancy in one out of its seven SDSS spectra.

\subsection{Upper limits to the orbital periods}
\label{s-porb}
The radial velocities of the secondary stars follow from Kepler's 3rd
law and depend on the stellar masses, the orbital period, and are
subject to geometric foreshortening by a factor $\sin i$, with $i$ the
binary inclination with regards to the line of sight:

\begin{equation}
\frac{(M_\mathrm{wd}\sin i)^3}{(M_\mathrm{wd}+M_\mathrm{sec})^2}
=\frac{\Porb K_\mathrm{sec}^3}{2\pi G}
\end{equation}

\noindent
with $K_\mathrm{sec}$ the radial velocity amplitude of the secondary
star, and $G$ the gravitational constant. This can be rearranged to
solve for the orbital period,

\begin{equation}
\Porb=\frac{2\pi G(M_\mathrm{wd}\sin i)^3}
   {(M_\mathrm{wd}+M_\mathrm{sec})^2K_\mathrm{sec}^3}
\label{e-porb}
\end{equation}

\noindent
From this equation, it is clear that assuming $i=90^\circ$ gives an
upper limit to the orbital period. 

The radial velocity measurements of our PCEBs and PCEB candidates
(Table\,\ref{t-rv}) sample the motion of their companion stars at
random orbital phases. However, if we \textit{assume} that the maximum
and minimum values of the observed radial velocities sample the
quadrature phases, e.g. the instants of maximum radial velocity, we
obtain \textit{lower limits} to the true radial velocity amplitudes of
the companion stars in our systems. From Eq.\,\ref{e-porb}, a lower
limit to $K_\mathrm{sec}$ turns into an upper limit to \Porb.

\begin{figure*}
\includegraphics[angle=-90,width=\textwidth]{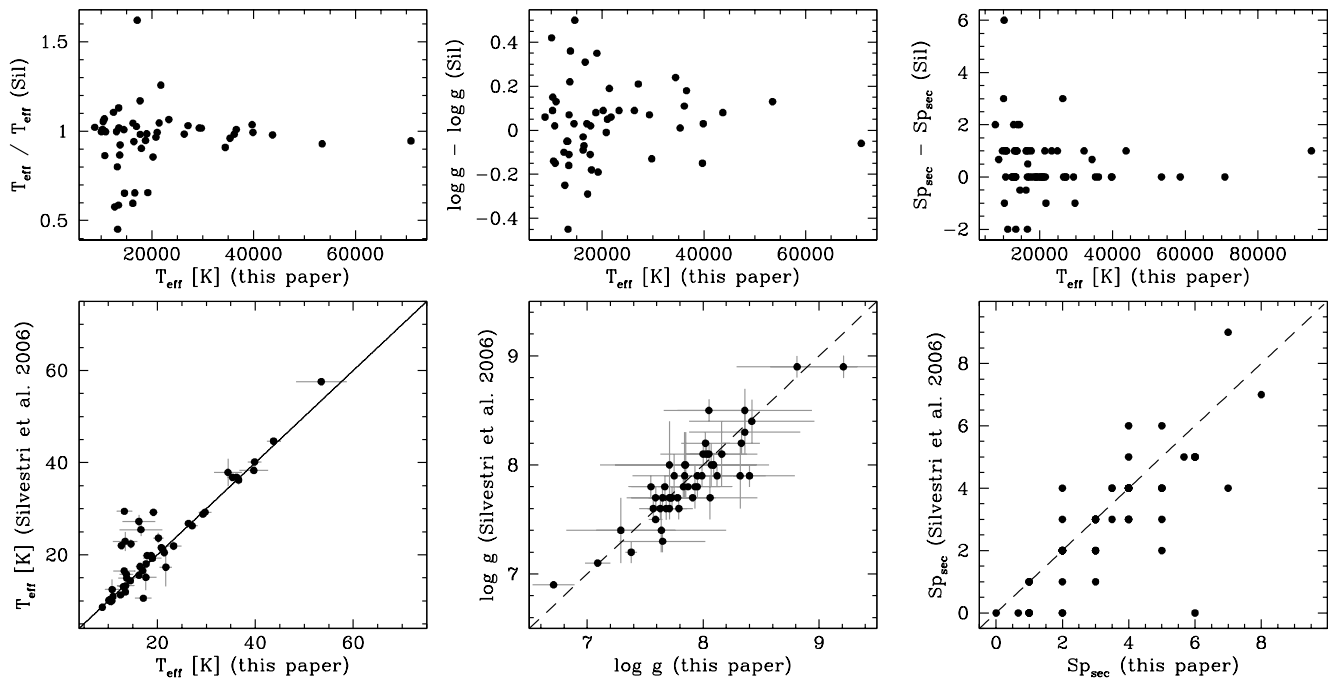}
\caption{\label{f-comp_sil}Comparison of the white dwarf effective 
temperatures and surface gravities and the spectral types of the secondary 
stars determined from our fits (Sect.\,\ref{s-decomposition}, \ref{s-wd} and
Table\,\ref{t-pcebfit}), and those of \citet{silvestrietal06-1}. Top
panels, from left to right: the WD effective temperature and surface gravity ratios,
and the difference in the secondary's spectral types from
the two studies as a function of the white dwarf temperature.}
\end{figure*}

Hence, combining the radial velocity information from
Table\,\ref{t-rv} with the stellar parameters from
Table\,\ref{t-pcebfit}, we determined upper limits to the orbital
periods of all PCEBs and PCEB candidates, which range between
0.46--7880\,d. The actual periods are likely to be substantially
shorter, especially for those systems where only two SDSS spectra are
available and the phase sampling is correspondingly poor. More
stringent constraints could be obtained from a more complex exercise
where the mid-exposure times are taken into account~--~however, given
the fact that many of the SDSS spectra are combined from data taken on
different nights, we refrained from this approach.

\subsection{The fraction of PCEB among the SDSS WDMS binaries} 
\label{s-frac}
We have measured the radial velocities of 101 WDMS which have multiple
SDSS spectra, and find that 15 of them clearly show radial
velocity variations, three additional WDMS are good candidates for
radial velocity variations (see Table\,\ref{t-rv}). Taking
the upper limits to the orbital periods at face value, and assuming
that systems with a period $\la300$\,d have undergone a CE
(\citealt{willems+kolb04-1}, see also Sect.\,\ref{s-PCEB}) 17
of the systems in Table\,\ref{t-rv} qualify as PCEBs, implying a PCEB
fraction of $\sim$15\,\% in our WDMS sample, which is
in rough agreement with the predictions by the population model of
\citet{willems+kolb04-1}. However, our value is likely to be a lower
limit on the true fraction of PCEBs among the SDSS WDMS binares for
the following reasons. (1) In most cases only two spectra are
available, with a non-negligible chance of sampling similar orbital
phases in both observations. (2) The relatively low spectral
resolution of the SDSS spectroscopy ($\lambda/ \Delta \lambda
\simeq1800$) plus the uncertainty in the flux calibration limit the
detection of significant radial velocity changes to $\sim
15\,\mathrm{km\,s^{-1}}$ for the best spectra. (3) In binaries with
extremely short orbital periods the long exposures will smear the
\Ion{Na}{I} doublet beyond recognition. (4) A substantial number of
the SDSS spectra are combined, averaging different orbital phases and
reducing the sensitivity to radial velocity changes. Follow-up
observations of a representative sample of SDSS WDMS  with
higher spectral resolution and a better defined cadence will be
necessary for an accurate determination of the fraction of PCEBs.

\subsection{Comparison with \citet{raymondetal03-1}}
\label{s-raym}
In a previous study, \citet{raymondetal03-1} determined white dwarf
temperatures, distance estimates based on the white dwarf fits, and
spectral types of the companion star for 109 SDSS WDMS. They
restricted their white dwarf fits to a single gravity, $\log g=8.0$,
and a white dwarf radius of $8 \times 10^{8}$\,cm (corresponding to
$M_\mathrm{wd}=0.6$\,M$_\odot$), which is a fair match for the
majority of systems (see Sect.\,\ref{s-distr} below). Our sample of
WDMS with two or more SDSS spectra has 28 objects in common with
Raymond's list, sufficient to allow for a quantitative comparison
between the two different methods used to fit the data. As we fitted
two or more spectra for each WDMS, we averaged for this purpose the
parameters obtained from the fits to individual spectra of a given
object, and propagated their errors accordingly. We find that
$\sim2/3$ of the temperatures determined by \citet{raymondetal03-1}
agree with ours at the $\sim20$ per cent level, with the remaining
being different by up to a factor two (Fig.\,\ref{f-comp_ray}, left
panels). This fairly large disagreement is most likely caused by the
simplified fitting Raymond et al. adopted, i.e. fitting the white
dwarf models in the wavelength range 3800--5000\,\AA, neglecting the
contribution of the companion star. The spectral types of the
companion stars from our work and \citet{raymondetal03-1} agree mostly
to within $\pm1.5$ spectral classes, which is satisfying given the
composite nature of the WDMS spectra and the problems associated with
their spectral decomposition (Fig.\,\ref{f-comp_ray}, right
panels). The biggest discrepancy shows up in the distances, with the
Raymond et al. distances being systematically lower than ours
(Fig.\,\ref{f-comp_ray}, middle panels). The average of the factor by
which Raymond et al. underpredict the distances is 6.5, which is close
to $2\pi$, suggesting that the authors may have misinterpreted the
flux definition of the model atmosphere code they used (TLUSTY/SYNSPEC
from \citealt{hubeny+lanz95-1}, which outputs Eddington fluxes), and
hence may have used a wrong constant in the flux normalisation
(Eq.\,\ref{e-wddist}).

\subsection{Comparison with \citet{silvestrietal06-1}}
\label{s-silves}

Having developed an independent method of determining the stellar
parameters for WDMS from their SDSS spectra, we compared our results
to those of \citet{silvestrietal06-1}. As in Sect.\,\ref{s-raym} above,
we average the parameters obtained from the fits to the individual SDSS
spectra of a given object. 
Figure\,\ref{f-comp_sil} shows
the comparison between the white dwarf effective temperatures, surface
gravities, and spectral types of the secondary stars from the two
studies. Both studies agree in broad terms for all three fit
parameters (Fig.\,\ref{f-comp_sil}, bottom panels). Inspecting the
discrepancies between the two independent sets of stellar parameters,
it became evident that relatively large disagreements are most
noticeably found for $\Teff\la20\,000$\,K, with differences in \Teff\
of up to a factor two, an order of magnitude in surface gravity, and a
typical difference in spectral type of the secondary of $\pm2$
spectral classes. For higher temperatures the differences become 
small, with nearly identical values for \Teff, $\log g$ agreeing
within $\pm0.2$\,magnitude, and spectral types differing by $\pm1$
spectral classes at most (Fig.\,\ref{f-comp_sil}, top panels). We
interpret this strong disagreement at low to intermediate white dwarf
temperatures to the ambiguity between hot and cold solutions described
in Sect.\,\ref{s-wd}. 

A quantitative judgement of the fits in \citet{silvestrietal06-1} is
difficult, as the authors do not provide much detail on the method
used to decompose the WDMS spectra, except for a single example in their
Fig.\,1. It is worth noting that the M dwarf component in that figure
displays constant flux at $\lambda<6000$\,\AA, which seems rather
unrealistic for the claimed spectral type of M5. Unfortunately,
\citet{silvestrietal06-1} do not list distances implied by their fits
to the white dwarf and main sequence components in their WDMS sample,
which would provide a test of internal consistency (see
Sect.\,\ref{s-distdist}).

We also investigated the systems \citeauthor{silvestrietal06-1}'s
(2006) method failed to fit, and found that we were able to determine
reasonable parameters for the majority of them. It appears that our
method is more robust in cases of low signal-to-noise ratio, and in
cases where one of the stellar components contributes relatively
little to the total flux. Examples of the latter are
SDSS\,J204431.45--061440.2, where an M0 secondary star dominates the
SDSS spectrum at $\lambda\ga4600$\,\AA, or SDSS\,J172406.14+562003.1,
which is a close PCEB containing a hot white dwarf and a low-mass
companion. An independent analysis of the entire WDMS sample from SDSS
appears therefore a worthwhile exercise, which we will pursue
elsewhere.

\subsection{Distribution of the stellar parameters}
\label{s-distr}

\begin{figure}
\includegraphics[width=0.49\textwidth]{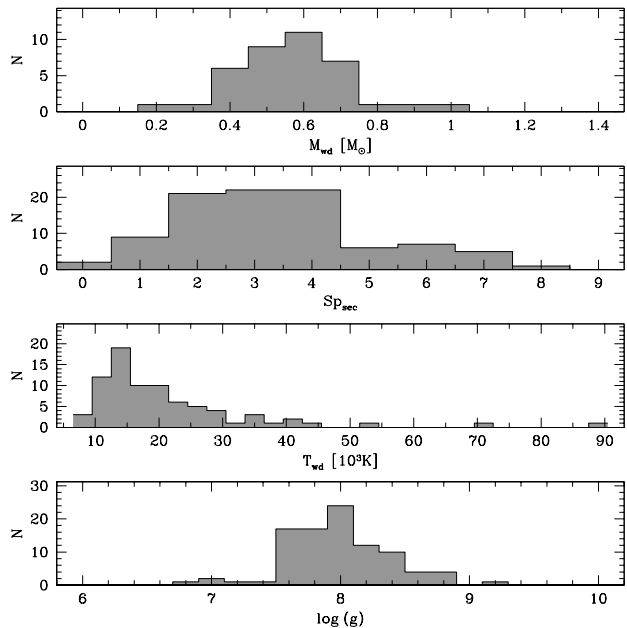}
\caption{\label{f-histo} White dwarf mass, Sp types of the secondaries,
effective temperature and $\log g$ histograms obtained from the SDSS WDSS 
sample. Excluded are those systems with individual WD masses, 
\Teff, and $\log g$ associated to relative errors larger than 25 per cent.}
\end{figure}

Having determined stellar parameters for each individual system in
Sect.\,\ref{s-stellar}, we are looking here at their global
distribution within our sample of WDMS.  Figure.\,\ref{f-histo} shows
histograms of the white dwarf effective temperatures, masses, $\log
g$, and the spectral types of the main-sequence companions. 

As in Sect.\,\ref{s-raym} and \ref{s-silves} above, we use here the
average of the fit parameters obtained from the different SDSS spectra
of each object. Furthermore, we exclude all systems with relative
errors in their white dwarf parameters ($T_\mathrm{wd}, \log g,
M_\mathrm{wd}$) exceeding 25 per cent to prevent smearing of the histograms
due to poor quality data and/or fits, which results in 95, 81, 94, and
38 WDMS in the histograms for the companion spectral type, $\log
g$, $T_\mathrm{wd}$, and $M_\mathrm{wd}$, respectively.  In broad
terms, our results are consistent with those of
\citet{raymondetal03-1} and \citet{silvestrietal06-1}: the most
frequent white dwarf temperatures are between 10\,000--20\,000\,K,
white dwarf masses cluster around $M_\mathrm{wd}\simeq0.6$\,M$_\odot$, and
the companion stars have most typically a spectral type M3--4, with
spectral types later than M7 or earlier than M1 being very rare.

At closer inspection, the distribution of white dwarf masses in our
sample has a more pronounced tail towards lower masses compared to the
distribution in \citet{silvestrietal06-1}. A tail of lower-mass white
dwarfs, peaking around 0.4\,M$_\odot$ is observed also in well-studied
samples of single white dwarfs \citep[e.g.][]{liebertetal05-1}, and is
interpreted as He-core white dwarfs descending from evolution in a
binary star \citep[e.g.][]{marshetal95-1}. In a sample of WDMS, a
significant fraction of systems will have undergone a CE phase, and
hence the fraction of He-core white dwarfs among WDMS is expected to
be larger than in a sample of single white dwarfs.

Also worth noting is that our distribution of companion star spectral
types is relatively flat between M2--M4, more similar to the
distribution of single M-dwarfs in SDSS \citep{westetal04-1} than the
companion stars in \citet{silvestrietal06-1}. More generally speaking,
the cut-off at early spectral types is due to the fact that WDMS with
K-type companions can only be identified from their spectra/colours if
the white dwarf is very hot~--~and hence, very young, and
correspondingly only few of such systems are in the total SDSS WDMS
sample. The cut-off seen for low-mass companions is not so trivial to
interpret. Obviously, very late-type stars are dim and will be harder
to be detected against a moderately hot white dwarf, such a bias was
discussed by \citet{schreiber+gaensicke03-1} for a sample of $\sim30$
well-studied WDMS which predominantly originated from blue-colour
(\,=\,hot white dwarf) surveys. However, old WDMS with cool white
dwarfs should be much more common \citep{schreiber+gaensicke03-1}, and
SDSS, sampling a much broader colour space than previous surveys,
should be able to identify WDMS containing cool white dwarfs plus very
late type companions. The relatively low frequency of such systems in
the SDSS spectroscopic data base suggests that either SDSS is not
efficiently targeting those systems for spectroscopic follow-up, or
that they are rare in the first place, or a combination of both. A
detailed discussion is beyond the scope of this paper, but we note
that \citet{farihietal05-1} have constructed the relative distribution
of spectral types in the local M/L dwarf distribution, which peaks
around M3--4, and steeply declines towards later spectral types,
suggesting that late-type companions to white dwarfs are intrinsically
rare. This is supported independently by
\citet{grether+lineweaver06-1}, who analysed the mass function of
companions to solar-like stars, and found that it steeply decreases
towards the late end of the main sequence (but rises again for
planet-mass companions, resulting in the term "brown dwarf desert").

An assessment of the stellar parameters of all WDMS in SDSS DR5 using
our spectral decomposition and white dwarf fitting method will 
improve the statistics of the distributions presented here, and will
be presented in a future paper. 

\begin{figure*}
\includegraphics[angle=-90,width=0.49\textwidth]{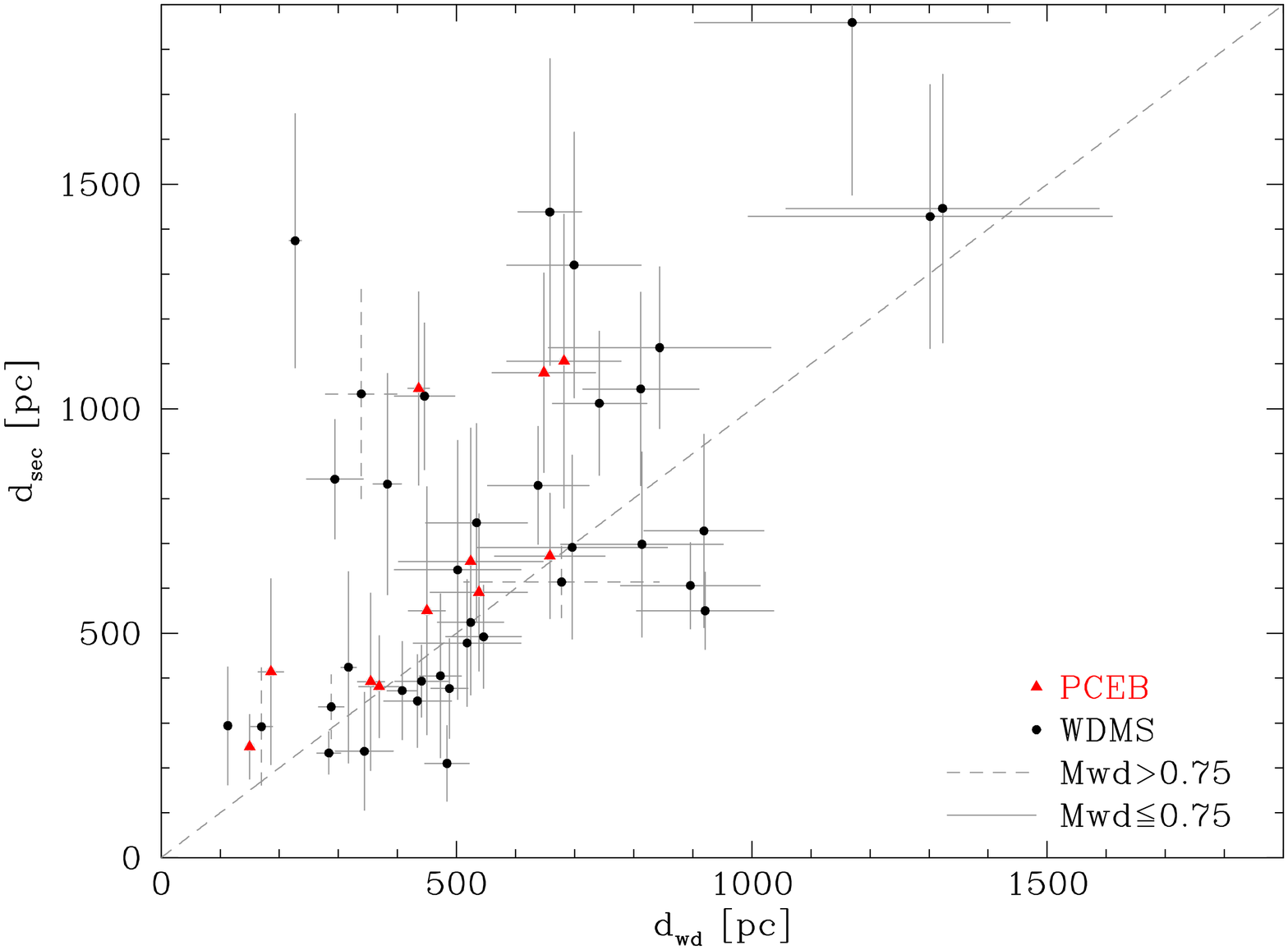}
\includegraphics[angle=-90,width=0.49\textwidth]{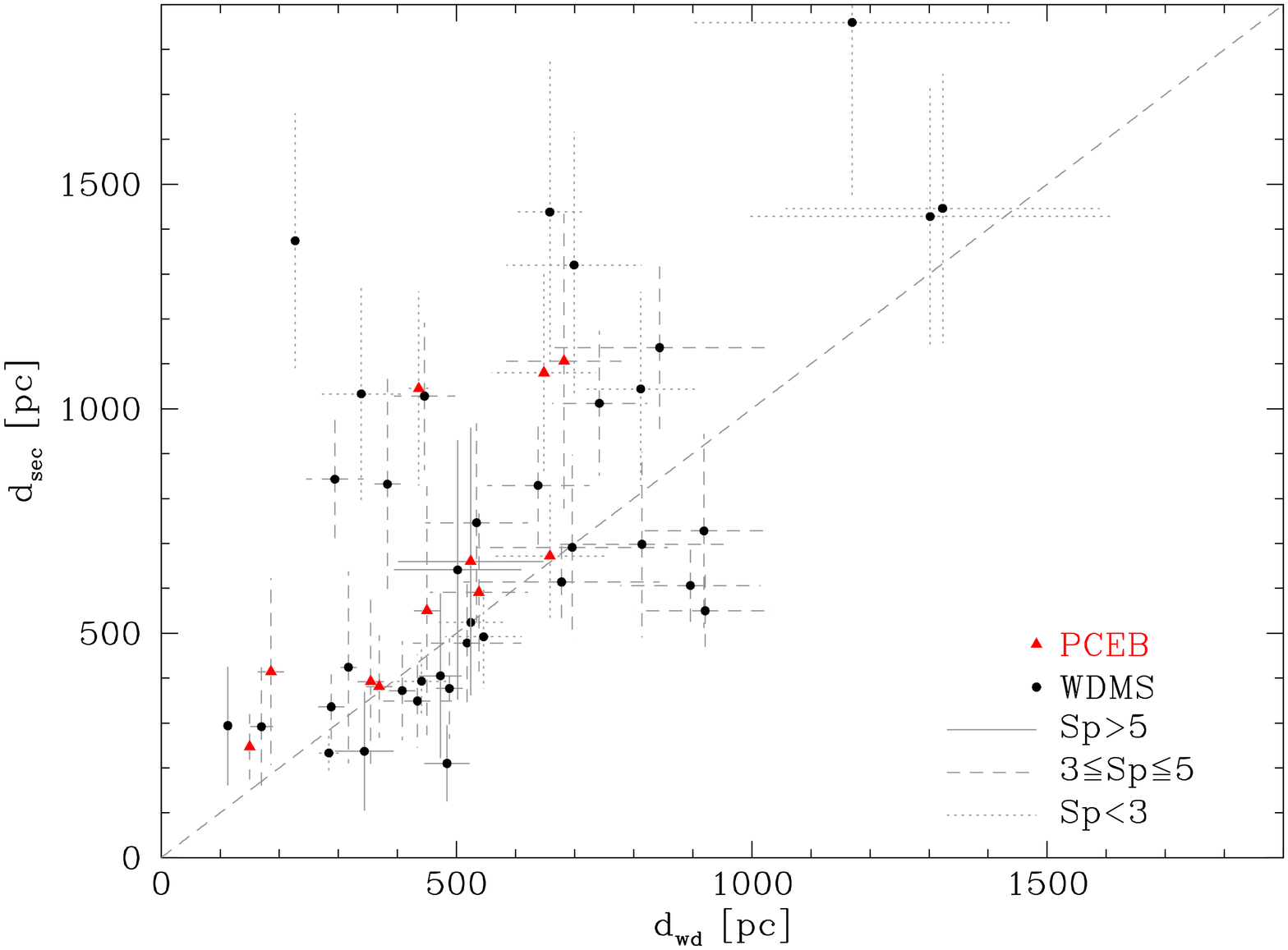}
\caption{\label{f-dd} Comparison of $d_\mathrm{sec}$ and
$d_\mathrm{wd}$ obtained from our spectral decomposition and white
dwarf fits to the SDSS spectra. Approximately a third of the systems
have $d_\mathrm{sec}\neq d_\mathrm{wd}$. The left panel splits the
sample according to the mass of the white dwarfs, while the right
panel divides the sample according to the spectral types of the
secondaries. In both panels systems that we identify as PCEBs from
radial velocity variations in their SDSS spectra are shown in red.}

\includegraphics[angle=-90,width=0.49\textwidth]{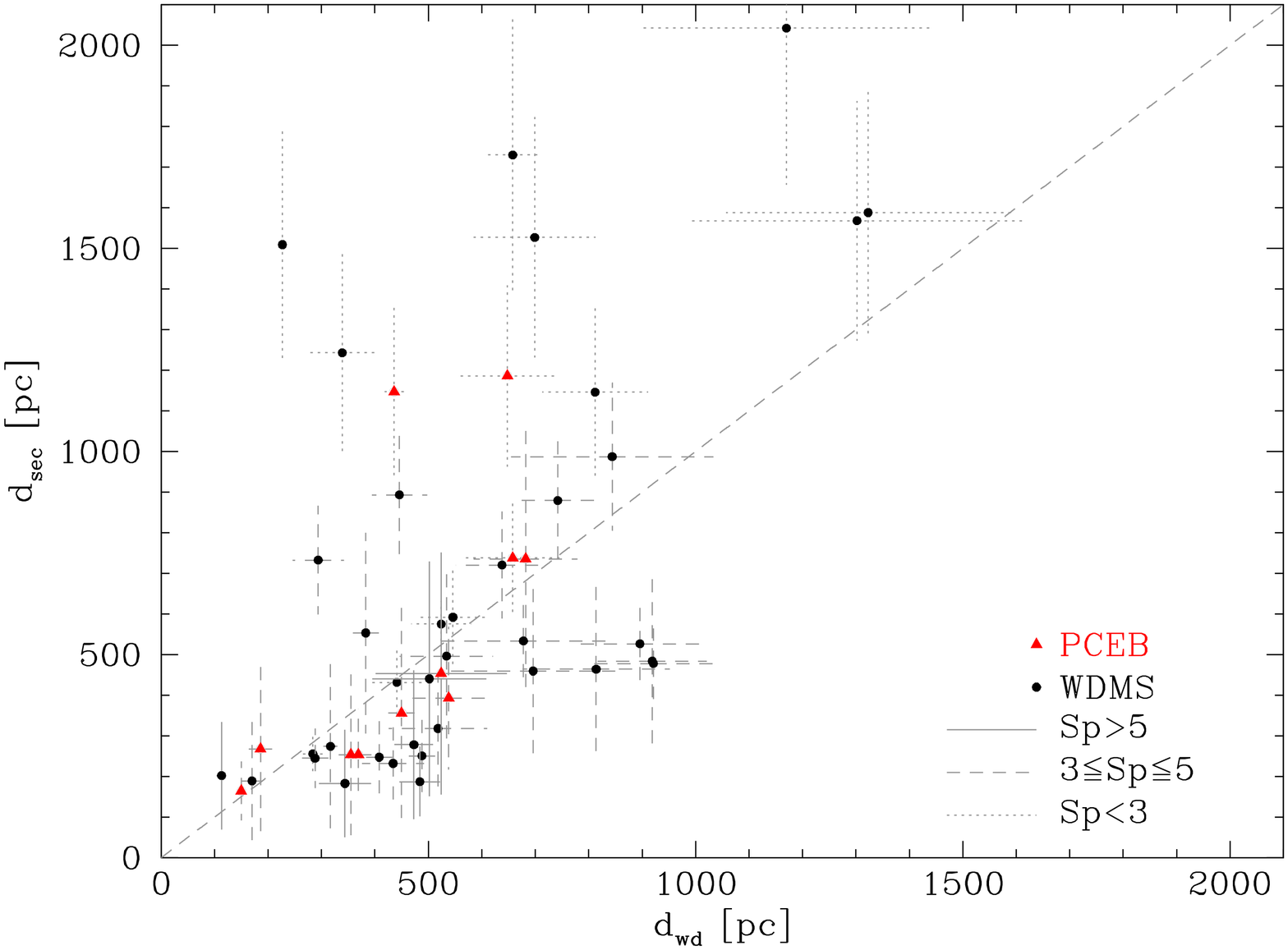}
\includegraphics[angle=-90,width=0.49\textwidth]{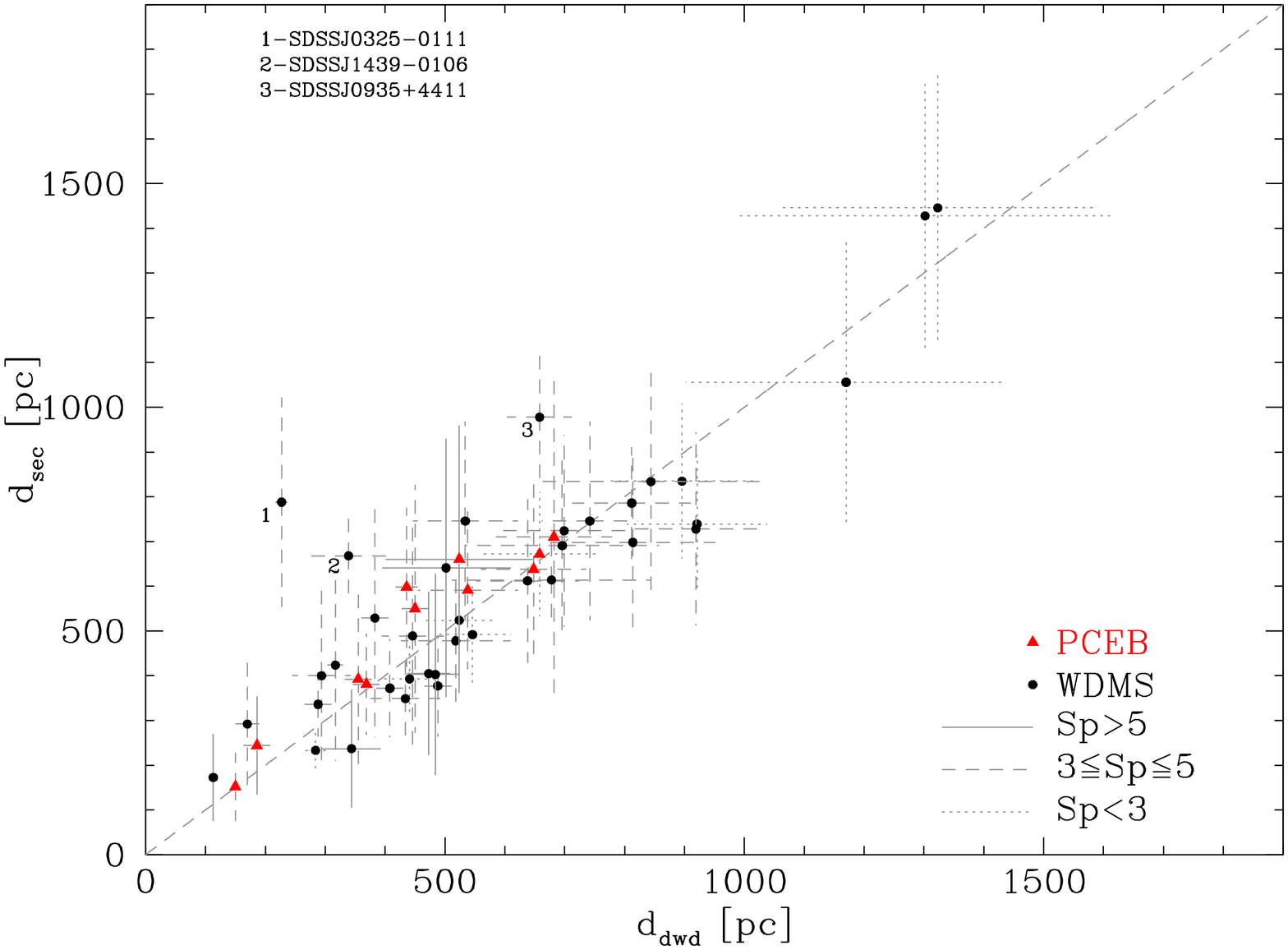}
\caption{\label{f-ddnew} Left panel: the distances implied by the
  spectral decomposition were calculated by using the $Sp-R$ relation predicted
  by the models of \citet{baraffeetal98-1}, instead of our empirical
  $Sp-R$ relation. Right panel: the spectral types of the secondary
  stars were adjusted by 1--2 spectral classes to achieve
  $d_\mathrm{wd}=d_\mathrm{sec}$. Only three systems can not be
  reconciled in this way, and are discussed individually in the
  text. We suggest that stellar activity in some WDMS may change the
  spectral type of their secondary stars, being equivalent to a change
  in surface temperature by a few 100\,K.}
\end{figure*}

\subsection{Stellar activity on the secondary stars?}
\label{s-distdist}

As outlined in Sect.\,\ref{s-distances}, the scaling factors used in
the modelling of the two spectral components of each WDMS provide two
independent estimates of the distance to the system. In principle,
both estimates should agree within their errors. Figure\,\ref{f-dd}
compares the white dwarf and secondary star distance estimates
obtained in Sect.\,\ref{s-distances}, where the distances obtained
from the individual SDSS spectra of a given object were averaged, and
the errors accordingly propagated. In this plot, we exclude systems
with relative errors in $d_\mathrm{wd}$ larger than 25 per cent to avoid
cluttering by poor S/N data.  The relative error in $d_\mathrm{sec}$
is dominated by the scatter in the $Sp-R$ relation, which represents an
intrinsic uncertainty rather than a statistical error in the fit, and
we therefore did not apply any cut in $d_\mathrm{sec}$. Taking the
distribution of distances at face value, it appears that about $2/3$
of the systems have $d_\mathrm{sec}\simeq d_\mathrm{wd}$ within their
1$\sigma$ errors, as expected from purely statistical
errors. However, there is a clear trend for outliers where
$d_\mathrm{sec}>d_\mathrm{wd}$. We will discuss the possible causes and
implications in the following sections.

\subsubsection{Possible causes for $d_\mathrm{sec}\neq d_\mathrm{wd}$}
\label{s-posscause}
We identify a number of possible causes for the discrepancy between
the two independent distance estimates observed in $\sim1/3$ of the
WDMS analysed here.

(1)~\textit{A tendency for systematic problems in the white dwarf
 fits?}  $d_\mathrm{sec}>d_\mathrm{wd}$ could be a result of too
small white dwarf radii for a number of systems, i.e. too high white
dwarf masses. We therefore identify in the left panel of
Fig.\,\ref{f-dd} those systems with massive $(>0.75$\,M$_\odot)$ white
dwarfs. It is apparent that the outliers from the
$d_\mathrm{sec}=d_\mathrm{wd}$ relation do not contain a large number
of very massive white dwarfs.

(2)~\textit{Problems in determining the correct spectral type of the
secondary?} If the error on the spectral type of the companion star
determined from the spectral decomposition is larger than $\pm0.5$, as
assumed in Sect.~\ref{s-decomposition}, a substantial deviation from
$d_\mathrm{sec}=d_\mathrm{wd}$ would result. However, as long as this
error is symmetric around the true spectral type, it would cause scatter
on both sides of the $d_\mathrm{sec}=d_\mathrm{wd}$ relation. 
Only if the determined spectral types were consistently too
early for $\sim1/3$ of the systems, the observed preference for outliers
at $d_\mathrm{sec}>d_\mathrm{wd}$ could be explained (see
Sect.~\ref{s-magneticeffects} below for a hypothetical \textit{systematic}
reason for spectral types that are consistently too early).

(3)~\textit{Problems in the spectral type-radius relation?} As
discussed in Sect.\,\ref{s-sp_r}, the spectral type-radius relation of
late type stars is not particularly well defined. The large scatter of
observed radii at a given spectral type is taken into account in the
errors in $d_\mathrm{sec}$. If those errors were underestimated,
they should cause an approximatively symmetric scatter of systems
around $d_\mathrm{sec}=d_\mathrm{wd}$, which is not observed (the $Sp-R$
relation being non-linear lead to asymmetric error bars in the radius
for a given symmetric error in the spectral type, however, over a
reasonably small range in the spectral type this effect is
negligible). A systematic problem over a small range of spectral types
would result in a concentration of the affected spectral types among
the outliers. For this purpose, we divide our sample into three groups
of secondary star spectral classes, Sp~$>$~5, 3~$\leq$~Sp $\leq$~5,
and Sp~$<$~3 (Fig.\,\ref{f-dd}, right panel). The outliers show a
slight concentration towards early types (Sp~$<$~3) compared to the
distribution of secondary star spectral types in the total sample
(Fig.\,~\ref{f-histo}).

To explore the idea that our empirical $Sp-R$ relation is simply inadequate, 
we calculated a new set of secondary star distances, using the
theoretical $Sp-R$ relation from \citet{baraffeetal98-1} (see
Fig.\,\ref{f-SpR}, bottom panel), which are shown in the left panel of
Fig.\,\ref{f-ddnew}. The theoretical $Sp-R$ relation implies smaller
radii in the range M3--M6, but the difference with our empirical
relation is not sufficient enough to shift the outlying WDMS onto the
$d_\mathrm{sec}=d_\mathrm{wd}$ relation. For spectral types earlier
than M2.5, our empirical $Sp-R$ relation actually gives \textit{smaller}
radii than the theoretical Baraffe et al. \citeyear{baraffeetal98-1}
relation, so that using the theoretical $Sp-R$ actually exacerbates the
$d_\mathrm{sec}>d_\mathrm{wd}$ problem.

(4)~\textit{A relationship with close binarity?} The fraction of PCEBs
among the outliers is similar to the fraction among the total sample
of WDMS (Fig.\,\ref{f-dd}), hence it does not appear that close
binarity is a decisive issue.

(5)~\textit{An age effect?} Late type stars take a long time to
contract to their zero age main sequence (ZAMS) radii, and if some of
the WDMS in our sample were relatively young objects, their M-dwarfs
would tend to have larger radii than ZAMS radii. As briefly discussed
in Footnote\,\ref{n-age}, the majority of the WDMS in our sample are
likely to be older than $\sim1$\,Gyr, and the outliers in
Fig.\,~\ref{f-dd},\ref{f-ddnew} do not show any preference for hot or
massive white dwarfs, which would imply short cooling ages and main
sequence life times.

\subsubsection{Could stellar activity affect Sp$_\mathrm{sec}$?}
\label{s-magneticeffects}

None of the points discussed in the previous section conclusively
explains the preference for outliers having
$d_\mathrm{sec}>d_\mathrm{wd}$. If we assume that the problem rests in
the determined properties of the secondary star, rather than those of
the white dwarf, the immediate implication of
$d_\mathrm{sec}>d_\mathrm{wd}$ is that the assumed radii of the
secondary stars are too large. As mentioned above and shown in
Fig.\,~\ref{f-ddnew}, this statement does not strongly depend on which
$Sp-R$ relation we use to determine the radii, either our empirical relation or
the theoretical Baraffe et al. (\citeyear{baraffeetal98-1}) relation. 
Rather than blaming the radii, we explore here whether the secondary
star spectral types determined from our decomposition of the SDSS
spectra might be consistently too early in the outlying systems.  If
this was the case, we would pick a radius from our $Sp-R$ relation that
is larger than the true radius of the secondary star, resulting in too
large a distance. In other words, the question is: \textit{is there a
mechanism that could cause the spectral type of an M-star, as derived
from low-resolution optical spectroscopy, to appear too early?} 

The reaction of stars to stellar activity on their surface, also
referred to as \textit{spottedness} is a complex phenomenon that is
not fully understood. Theoretical studies
\citep[e.g.][]{spruit+weiss86-1, mullan+macdonald01-1,
chabrieretal07-1} agree broadly on the following points: (1) the effect of
stellar activity is relatively weak at the low-mass end of the main
sequence ($M\la0.3$\,M$_\odot$), where stars are conventionally thought
to become fully convective (though, see \citealt{mullan+macdonald01-1,
chabrieretal07-1} for discussions on how magnetic fields may change
that mass boundary), (2) stellar activity will result in an increase
in radius, and (3) the effective temperature of an active star is
lower than that of an unspotted star. 

Here, we briefly discuss the possible effects of stellar activity on
the spectral type of a star. For this purpose, it is important not to
confuse the \textit{effective temperature}, which is purely a
definition coupled to the luminosity and the stellar radius via
$L=4\pi R^2 \sigma T_\mathrm{eff}^4$ (and hence is a \textit{global}
property of the star), and the \textit{local} temperature of a given
part of the stellar surface, which will vary from spotted areas to
inter-spot areas. In an unspotted star effective and local temperature
are the same, and both colour and spectral type are well-defined.  As
a simple example to illustrate the difference between effective
temperature and colour in an active star, we assume that a large
fraction of the star is covered by zero-temperature, i.e. black
spots, and that the inter-spot temperature is the same as that of the
unspotted star. As shown by \citet{chabrieretal07-1}, assuming constant
luminosity requires the radius of the star to increase, and the
effective temperature to drop. Thus, while intuition would suggest
that a lower effective temperature would result in a redder colour,
this ficticious star has \textit{exactly} the same colour and spectral
type as its unspotted equivalent~--~as the black spots contribute no
flux at all, and the inter-spot regions with the same spectral shape
as the unspotted star. 

Obviously, the situation in a real star will be more complicated, as
the spots will not be black, but have a finite temperature, and the
star will hence have a complicated temperature distribution over its
surface. Thus, the spectral energy distribution of such a spotted star
will be the superposition of contributions of different temperatures,
weighted by their respective covering fraction of the stellar surface.
Strictly speaking, such a star has no longer a well-defined spectral
type or colour, as these properties will depend on the wavelength
range that is observed. \citet{spruit+weiss86-1} assessed the effect
of long-term spottedness on the temperature distribution on active
stars, and found that for stars with masses in the range
$0.3-0.6$\,M$_\odot$ the long-term effect of spots is to increase the
temperature of the inter-spot regions by $\sim100-200$\,K (compared to
the effective temperature of the equivalent unspotted star), wheras
the inter-spot temperature of spotted lower-mass stars remains
unchanged. \citet{spruit+weiss86-1} also estimated the effects of
stellar activity on the colours of stars, but given their use of
simple blackbody spectra, these estimates are of limited value. As a
general tendency, the hotter (unspotted) parts of the star will
predominantly contribute in the blue end of the the spectral energy
distribution, the cooler (spotted) ones in its red end. As we
determine the spectral types of the secondary stars in the SDSS WDMS
from optical (\,=\,blue) spectra, and taking the results of
\citeauthor{spruit+weiss86-1} at face value, it appears hence possible
that they are too early compared to unspotted stars of the same
mass. A full theoretical treatment of this problem would involve
calculating the detailed surface structure of active stars as well as
appropriate spectral models for each surface element in order to
compute the spatially integreated spectrum as it would be observed. This is
clearly a challenging task.

Given that theoretical models on the effect of stellar activity have
not yet converged, and are far from making detailed predictions on the
spectroscopic appearence of active stars, we pursue here an empirical
approach. We assume that the discrepancy
$d_\mathrm{sec}>d_\mathrm{wd}$ results from picking a spectral type
too early, i.e. we assume that the secondary star appears hotter in
the optical spectrum that it should for its given mass. Then, we check
by how much we have to adjust the spectral type (and the corresponding
radius) to achieve $d_\mathrm{sec}=d_\mathrm{wd}$ within the
errors. We find that the majority of systems need a change of 1--2
spectral classes, which corresponds to changes in the effective
temperature of a few hundred degrees only, in line with the
calculations of \citet{spruit+weiss86-1}. Bearing in mind that what we
\textit{see} in the optical is the surface temperature, and not the
effective temperature, comparing this to the surface temperature
changes calculated by \citet{spruit+weiss86-1}, and taking into
account that we ignored in this simple approach the change in radius
caused by a large spottedness, it appears plausible that the large
deviations from $d_\mathrm{sec}=d_\mathrm{wd}$ may be related to
stellar activity on the secondary stars. 

There are three WDMS where a change of more than two spectral classes
would be necessary: SDSSJ032510.84-011114.1, SDSSJ093506.92+441107.0,
and SDSSJ143947.62-010606.9. SDSSJ143947.62-010606.9 contains a very
hot white dwarf, and the secondary star may be heated if this system
is a PCEB. Its two SDSS spectra reveal no significant radial velocity
variation, but as discussed in Sect.\,\ref{s-PCEB} the SDSS spectra can
not exclude a PCEB nature because of random phase sampling, low
inclination and limited spectral resolution.  SDSSJ032510.84-011114.1
and SDSSJ093506.92+441107.0 could be short-period PCEBs, as they both
have poorly define \Ion{Na}{I} absorption doublets, possibly smeared
by orbital motion over the SDSS exposure (see Sect.\,\ref{s-PCEB}). In a
close binary, their moderate white dwarf temperatures would be
sufficient to cause noticeable heating of the secondary star. 
We conclude that our study suggests some anomalies in the
properties of $\sim1/3$ of the M-dwarf companions within the WDMS
sample analysed here. This is in line with previous detailed studies
reporting the anomalous behaviour of the main sequence companions in
PCEBs and cataclysmic variables, e.g. \citet{obrienetal01-1} or
\citet{nayloretal05-1}.

\subsection{Selection effects among the SDSS WDMS.}
Selection effects among the WDMS found by SDSS with respect to
the spectral type of their main-sequence component can be deduced from
the right panel of Fig.\,\ref{f-dd}.  No binaries with econdary
spectral types later than M5 are found at distances larger than
$\sim500$\,pc. Because of their intrinsic faintness, such late-type
secondary stars can only be seen against relatively cool white dwarfs,
and hence the large absolute magnitude of such WDMS limits their
detection within the SDSS magnitude limit to a relatively short
distance. Hot white dwarfs in SDSS can be detected to larger
distances, and may have undetected late-type companions. There are
also very few WDMS with secondary stars earlier than M3 within
500\,pc. In those systems, the secondary star is so bright that it
saturates the $z$, and possibly the $i$ band, disqualifying the
systems for spectroscopic follow up by SDSS. While these selection
effects may seem dishearting at first, it will be possible to
quantitatively correct them based on predicted colours of WDMS
binaries and the information available within the SDSS project
regarding photometric properties and spectroscopic selection
algorithms.

\section{Conclusions}
We have identified 18 PCEBs and PCEB candidates among a sample of 101
WDMS for which repeat SDSS spectroscopic observations are available in
DR5. From the SDSS spectra, we determine the spectral types of the
main sequence companions, the effective temperatures, surface
gravities, and masses of the white dwarfs, as well as distance
estimates to the systems based both on the properties of the white
dwarfs and of the main sequence stars. In about 1/3 of the WDMS
studied here the SDSS spectra suggest that the secondary stars have
either radii that are substantially larger than those of single
M-dwarfs, or spectral types that are too early for their masses. 
Follow-up observations of the PCEBs and PCEB candidates is encouraged
in order to determine their orbital periods as well as more detailed
system parameters. Given the fact that we have analysed here only
$\sim10$ per cent of the WDMS in DR5, it is clear SDSS holds the potential to
dramatically improve our understanding of CE evolution.

\section*{Acknowledgements.}
ARM and BTG were supported by a PPARC-IAC studentship and by an
Advanced Fellowship, respectively. MRS ackowledges support by FONDECYT
(grant 1061199). We thank Isabelle Baraffe for providing the
theoretical spectral type-radius relation shown in Fig.\,\ref{f-SpR}
and for discussions on the properties of low mass stars, and
the referee Rob Jeffries for a fast and useful report. 

\label{lastpage}


\onecolumn
\setlength{\tabcolsep}{0.9ex}
\begin{longtable}{ccccccccccccccccc}
\caption{\label{t-pcebfit}WD masses, effective temperatures, surface gravities,
spectral types and distances of the SDSS PCEBs identified in Sect.\,\ref{s-stellar}, as 
determined from spectral modelling. The stellar parameters for the remaining 112 WDMS binaries
can be found in the electronic edition of the paper. We quote by and $s$ and $e$ those
systems which have been studied previously by \citet{silvestrietal06-1} and 
\citet{eisensteinetal06-1}, repectively.}\\[-1.0ex]
\hline\noalign{\smallskip}
%
SDSS\,J		        &	MJD  &	plate & fiber	&	      \Teff(K)	&       err	&    $\log g$   &   err   &     $M(M_{\bigodot})$     &    err	&   $d_\mathrm{wd}$(pc)  &      err   & Sp  &	 $d_\mathrm{sec}$(pc)  &   err	   &    flag &   notes \\
\noalign{\smallskip}\hline\noalign{\smallskip}
\endfirsthead
%
\multicolumn{17}{c}{{\tablename} \thetable{} -- Continued}\\
\noalign{\smallskip}\hline\noalign{\smallskip}
SDSS\,J		        &	MJD  &	plate & fiber	&	      \Teff(K)	&       err	&    $\log g$   &   err   &     $M(M_{\bigodot})$     &    err	&   $d_\mathrm{wd}$(pc)  &      err   & Sp  &	 $d_\mathrm{sec}$(pc)  &   err	   &    flag &   notes \\
\noalign{\smallskip}\hline\noalign{\smallskip}\endhead
%
\noalign{\smallskip}\hline\noalign{\smallskip}
\multicolumn{17}{l}{{Continued on Next Page\ldots}} \endfoot
%
\noalign{\smallskip}\hline\noalign{\smallskip}
\multicolumn{17}{p{\textwidth}}{(1) \Teff\, less than 6000; (2) Noisy spectra; 
(3) Cold WD; (4) Diffuse background galaxy in the SDSS
image; (5) Reflection effect; (6) Some blue excess, WD?} \\
\endlastfoot
%
 005245.11-005337.2	&	51812  &   394 &    96	&    15071 &  4224 &   8.69 &  0.73 &  1.04 &  0.38 &   505 &   297  &   4 &    502 &   149 &   s,e 	  &  	  \\
 			&	51876  &   394 &    100	&    17505 &  7726 &   9.48 &  0.95 &  1.45 &  0.49 &   202 &    15  &   4 &    511 &   152 &   	  &	  \\
 			&	51913  &   394 &    100	&    16910 &  2562 &   9.30 &  0.42 &  1.35 &  0.22 &   261 &   173  &   4 &    496 &   147 &   	  &	  \\
 			&	52201  &   692 &    211	&    17106 &  3034 &   9.36 &  0.43 &  1.38 &  0.22 &   238 &   178  &   4 &    526 &   156 &   	  &	  \\
 005457.61-002517.0	&	51812  &   394 &    118	&    16717 &   574 &   7.81 &  0.13 &  0.51 &  0.07 &   455 &    38  &   5 &    539 &   271 &   s,e	  &	  \\
 			&	51876  &   394 &    109	&    17106 &   588 &   7.80 &  0.14 &  0.51 &  0.07 &   474 &    40  &   5 &    562 &   283 &  		  &	  \\
 			&	51913  &   394 &    110	&    17106 &   290 &   7.88 &  0.07 &  0.55 &  0.04 &   420 &    19  &   5 &    550 &   277 &  		  &	  \\
 022503.02+005456.2	&	51817  &   406 &    533	&        - &     - &      - &     - &     - &     - &     - &     -  &   5 &    341 &   172 &   s,e  	  & 1	  \\
 			&	51869  &   406 &    531	&        - &     - &      - &     - &     - &     - &     - &     -  &   5 &    351 &   177 &		  &	  \\
 			&	51876  &   406 &    532	&        - &     - &      - &     - &     - &     - &     - &     -  &   5 &    349 &   176 &		  &	  \\
 			&	51900  &   406 &    532	&        - &     - &      - &     - &     - &     - &     - &     -  &   5 &    342 &   172 &		  &	  \\
 			&	52238  &   406 &    533	&        - &     - &      - &     - &     - &     - &     - &     -  &   5 &    356 &   179 &		  &	  \\
 024642.55+004137.2	&	51871  &   409 &    425	&    15782 &  5260 &   9.18 &  0.76 &  1.29 &  0.39 &   213 &   212  &   4 &    365 &   108 &   s,e	  &	  \\
 			&	52177  &   707 &    460	&        - &     - &      - &     - &     - &     - &     - &     -  &   3 &    483 &    77 &		  &	  \\
 			&	52965  &   1664&    420	&    16717 &  1434 &   8.45 &  0.28 &  0.90 &  0.16 &   515 &   108  &   3 &    492 &    78 &		  &	  \\
 			&	52973  &   1664&    407	&    14065 &  1416 &   8.24 &  0.22 &  0.76 &  0.14 &   510 &    77  &   3 &    499 &    80 &		  &	  \\
 025147.85-000003.2	&	52175  &   708 &    228	&    17106 &  4720 &   7.75 &  0.92 &  0.49 &  0.54 &  1660 &   812  &   4 &    881 &   262 &   e	  & 2	  \\
 			&	52177  &   707 &    637	&        - &     - &      - &     - &     - &     - &     - &     -  &   4 &    794 &   236 &		  &	  \\
 030904.82-010100.8	&	51931  &   412 &    210	&    19416 &  3324 &   8.18 &  0.68 &  0.73 &  0.40 &  1107 &   471  &   3 &    888 &   141 &   s,e	  &	  \\
 			&	52203  &   710 &    214	&    18756 &  5558 &   9.07 &  0.61 &  1.24 &  0.31 &   462 &   325  &   3 &    830 &   132 &  		  &	  \\
 			&	52235  &   412 &    215	&    14899 &  9359 &   8.94 &  1.45 &  1.17 &  0.75 &   374 &   208  &   4 &    586 &   174 &  		  &	  \\
 			&	52250  &   412 &    215	&    11173 &  9148 &   8.55 &  1.60 &  0.95 &  0.84 &   398 &   341  &   4 &    569 &   169 &  		  &	  \\
 			&	52254  &   412 &    201	&    20566 &  7862 &   8.82 &  0.72 &  1.11 &  0.37 &   627 &   407  &   3 &    836 &   133 &  		  &	  \\
 			&	52258  &   412 &    215	&    19640 &  2587 &   8.70 &  0.53 &  1.04 &  0.27 &   650 &   281  &   3 &    854 &   136 &  		  &	  \\
 			&	53386  &   2068&     126&    15246 &  4434 &   8.75 &  0.79 &  1.07 &  0.41 &   522 &   348  &   4 &    628 &   187 &  		  &	  \\
 031404.98-011136.6	&	51931  &   412 &    45	&        - &     - &      - &     - &     - &     - &     - &     -  &   4 &    445 &   132 &   s,e  	  & 1	  \\
 			&	52202  &   711 &    285	&        - &     - &      - &     - &     - &     - &     - &     -  &   4 &    475 &   141 & 		  &	  \\
 			&	52235  &   412 &    8	&        - &     - &      - &     - &     - &     - &     - &     -  &   4 &    452 &   134 & 		  &	  \\
 			&	52250  &   412 &    2	&        - &     - &      - &     - &     - &     - &     - &     -  &   4 &    426 &   126 & 		  &	  \\
 			&	52254  &   412 &    8	&        - &     - &      - &     - &     - &     - &     - &     -  &   4 &    444 &   132 & 		  &	  \\
 			&	52258  &   412 &    54	&        - &     - &      - &     - &     - &     - &     - &     -  &   4 &    445 &   132 & 		  &	  \\
 082022.02+431411.0	&	51959  &   547 &    76	&    21045 &   225 &   7.94 &  0.04 &  0.59 &  0.02 &   153 &     4  &   4 &    250 &    74 &   s,e	  &	  \\
 			&	52207  &   547 &    59	&    21045 &   147 &   7.95 &  0.03 &  0.60 &  0.01 &   147 &     2  &   4 &    244 &    72 &   	  &	  \\
 113800.35-001144.4	&	51630  &   282 &    113	&    18756 &  1364 &   7.99 &  0.28 &  0.62 &  0.17 &   588 &   106  &   4 &    601 &   178 &   s,e  	  & 	  \\
 			&	51658  &   282 &    111	&    24726 &  1180 &   8.34 &  0.16 &  0.84 &  0.10 &   487 &    60  &   4 &    581 &   173 & 		  &	  \\
 115156.94-000725.4	&	51662  &   284 &    435	&    10427 &   193 &   7.90 &  0.23 &  0.54 &  0.14 &   180 &    25  &   5 &    397 &   200 &   s,e	  &	  \\
 			&	51943  &   284 &    440	&    10189 &   115 &   7.99 &  0.16 &  0.59 &  0.10 &   191 &    19  &   5 &    431 &   217 & 		  &	  \\
 152933.25+002031.2	&	51641  &   314 &    354	&    14228 &   575 &   7.67 &  0.12 &  0.44 &  0.05 &   338 &    25  &   5 &    394 &   199 &   s,e	  &	  \\
 			&	51989  &   363 &    350	&    14728 &   374 &   7.59 &  0.09 &  0.41 &  0.04 &   372 &    21  &   5 &    391 &   197 &  		  &	  \\
 172406.14+562003.0	&	51813  &   357 &    579	&    35740 &   187 &   7.41 &  0.04 &  0.42 &  0.01 &   417 &    15  &   2 &   1075 &   222 &   s,e	  &	  \\
 			&	51818  &   358 &    318	&    36154 &   352 &   7.33 &  0.06 &  0.40 &  0.02 &   453 &    24  &   2 &   1029 &   213 &   	  &	  \\
 			&	51997  &   367 &    564	&    37857 &   324 &   7.40 &  0.04 &  0.43 &  0.01 &   439 &    16  &   2 &   1031 &   213 &  		  &	  \\
 172601.54+560527.0	&	51813  &   357 &    547	&    20331 &  1245 &   8.24 &  0.23 &  0.77 &  0.14 &   582 &    94  &   2 &   1090 &   225 &   s,e	  &	  \\
 			&	51997  &   367 &    548	&    20098 &   930 &   7.94 &  0.18 &  0.59 &  0.11 &   714 &    83  &   2 &   1069 &   221 &   	  &	  \\
 173727.27+540352.2	&	51816  &   360 &    165	&    13127 &  1999 &   7.91 &  0.42 &  0.56 &  0.26 &   559 &   140  &   6 &    680 &   307 &   s,e	  &	  \\
 			&	51999  &   362 &    162	&    13904 &  1401 &   8.24 &  0.31 &  0.76 &  0.20 &   488 &   106  &   6 &    639 &   288 &  		  &	  \\
 224139.02+002710.9	&	53261  &   1901&    471	&    12681 &   495 &   8.05 &  0.15 &  0.64 &  0.09 &   369 &    35  &   4 &    381 &   113 &   e	  &	  \\
 			&	52201  &   674 &    625	&    13745 &  1644 &   7.66 &  0.36 &  0.43 &  0.19 &   524 &   108  &   4 &    378 &   112 &    	  &	  \\
 233928.35-002040.0	&	53357  &   1903&    264	&    15071 &  1858 &   8.69 &  0.33 &  1.04 &  0.18 &   416 &   112  &   4 &    530 &   157 &   e	  &	  \\
 			&	52525  &   682 &    159	&    12536 &  2530 &   7.92 &  0.79 &  0.56 &  0.48 &   655 &   291  &   4 &    528 &   157 &   	  &	  \\
 234534.49-001453.7	&	52524  &   683 &    166	&    19193 &  1484 &   7.79 &  0.31 &  0.51 &  0.17 &   713 &   132  &   4 &   1058 &   314 &   s,e 	  & 5	  \\
 			&	53357  &   1903&    103	&    18974 &   730 &   7.98 &  0.15 &  0.61 &  0.09 &   652 &    62  &   4 &   1155 &   343 &		  &	  \\
 235020.76-002339.9	&	51788  &   386 &    228	&        - &     - &      - &     - &     - &     - &     - &     -  &   5 &    504 &   254 &		  & 6	  \\
 			&	52523  &   684 &    226	&        - &     - &      - &     - &     - &     - &     - &     -  &   5 &    438 &    22 &		  &	  \\                                
\end{longtable}
\twocolumn

\end{document}